\documentclass[12pt,onecolumn]{IEEEtran}
\usepackage{setspace}

\usepackage{float}
\usepackage{amsmath,amsfonts}
\usepackage{algorithmic}
\usepackage{algorithm}
\usepackage{array}
\usepackage{ulem}\usepackage{ulem}
\usepackage{booktabs}
\usepackage[caption=false,font=footnotesize,labelfont=rm,textfont=rm]{subfig}
\usepackage{balance}
\usepackage{textcomp}
\usepackage{stfloats}
\usepackage{url}
\usepackage{ulem}
\usepackage{verbatim}
\usepackage{graphicx}
\usepackage{cite}
\usepackage{svg}
\usepackage{flushend}
\newcommand{\nop}[1]{}
\begin{document}

\title{Sensing and Storing Less: A MARL-based Solution for Energy Saving in Edge Internet of Things}
\author{{Zongyang Yuan, Lailong Luo,  Qianzhen Zhang, Bangbang Ren, Deke Guo, Richard T.B. Ma}
\thanks{ Zongyang Yuan, Qianzhen Zhang, Bangbang Ren, and Deke Guo are with Science and Technology on Information Systems Engineering Laboratory, National University of Defense Technology, Changsha, Hunan 410073, China. (e-mail: {yuanzongyang, zhangqianzhen18, renbangbang11}@nudt.edu.cn, {guodeke} @gmail.com).

Lailong Luo is with Science and Technology on Information Systems Engineering Laboratory, National University of Defense Technology, Changsha, Hunan 410073, China, and also with the National Laboratory for Parallel and Distributed Processing, National University of Defense Technology, Changsha, Hunan 410073, China. (e-mail: luolailong09@nudt.edu.cn).}

\thanks{Richard T. B. Ma is with the Department of Computer
Science, National University of Singapore, Singapore 119077 (e-mail:
tbma@comp.nus.edu.sg).}

\thanks{Corresponding author: Lailong Luo.} 
}

\maketitle

\begin{abstract}
\doublespacing  % 设置双倍行距
\normalem
As the number of Internet of Things (IoT) devices continuously grows and application scenarios constantly enrich, the volume of sensor data experiences an explosive increase. However, substantial data demands considerable energy during computation and transmission. Redundant deployment or mobile assistance is essential to cover the target area reliably with fault-prone sensors. Consequently, the ``butterfly effect" may appear during the IoT operation, since unreasonable data overlap could result in many duplicate data.
To this end, we propose \textit{Senses}, a novel online energy saving solution for edge IoT networks, with the insight of \underline{sen}sing and \underline{s}toring l\underline{es}s at the network edge by adopting Muti-Agent Reinforcement Learning (MARL). Senses achieves data de-duplication by dynamically adjusting sensor coverage at the sensor level. For exceptional cases where sensor coverage cannot be altered, Senses conducts data partitioning and eliminates redundant data at the controller level. Furthermore, at the global level, considering the heterogeneity of IoT devices, Senses balances the operational duration among the devices to prolong the overall operational duration of edge IoT networks. We evaluate the performance of Senses through testbed experiments and simulations. The results show that Senses saves $11.37\%$ of energy consumption on control devices and prolongs $20\%$ overall operational duration of the IoT device network.

\end{abstract}
\onehalfspacing  % 设置双倍行距


\section{Introduction}
\doublespacing  % 设置双倍行距

With the closer integration of artificial intelligence and $5$G technology, the application scenarios of the Internet of Things (IoT) have been continuously enriched~\cite{hospedales2021meta, kapidis2021multi, liu2022convnet}. Ubiquitous applications spawned intelligent devices with diversified functionalities~\cite{huang2022multimodal, tang2021reusing, yan2021scalable, 9658203}. In IoT networks, intelligent devices interact with each other to share and exchange data. Based on such interactions, data from different sensors are integrated or analyzed for diverse purposes and applications. The Global System for Mobile communication Association (GSMA) predicts that by 2025, the number of connected IoT devices worldwide (including cellular and non-cellular) will reach approximately $24.6$ billion~\cite{citation-key}, indicating a continuous and rapid growth trend in IoT connections.

\begin{figure}[t]%
\centering
\includegraphics[scale=0.7]{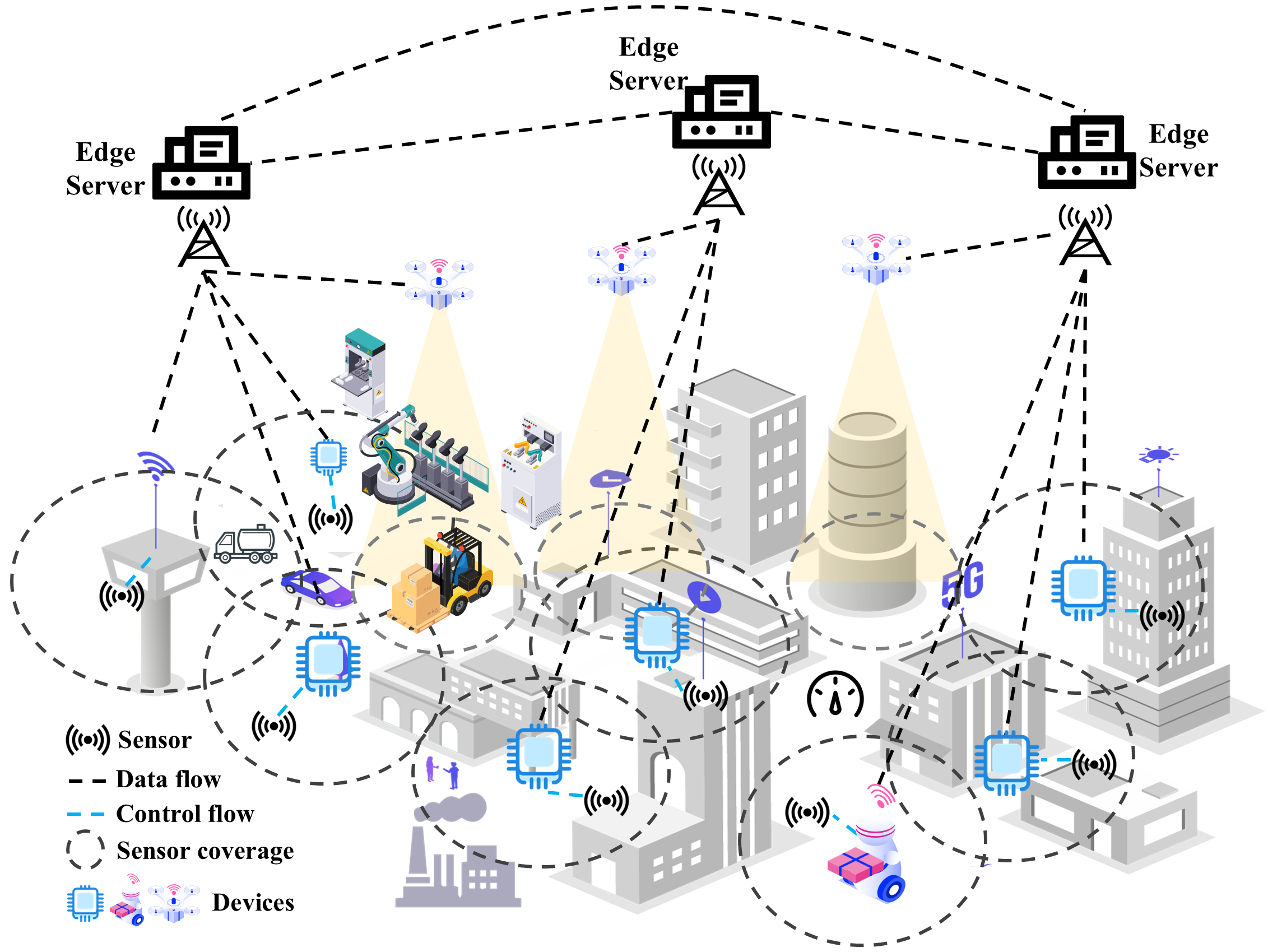}
\caption{An example of edge-assistant IoT networks.}%
\label{fig1}%
\vspace{-0.2in}
\end{figure}

The ever-increasing number of IoT devices and the continuous enrichment of application scenarios bring about explosive growth in the volume of data. This presents vast opportunities for data-hungry models and applications~\cite{yang2024drhouse,ge2024llm,liu2024llm}. However, a large number of data require a significant amount of energy during computation and transmission, thus arousing concerns about the energy consumption of the IoT~\cite{apiletti2011energy}. On the one hand, sensors are prone to malfunctions, such as those caused by electromagnetic interference. To ensure reliable coverage of the entire area, redundant deployment or mobile assistance is necessary~\cite{yu2018placement}. As a result, there is a high likelihood of a substantial amount of repetitive data in the IoT~\cite{afroze2021target, malek2016balanced}, which may give rise to a profound ``butterfly effect"\footnote{Meaning that duplicate data from data sources may accumulate a high energy overhead as the entire system transmits and computes chain reactions}. For IoT devices with energy shortages, it is unjustifiable to expend energy on transmitting and processing low-value repetitive data~\cite{jin2017q, han2019high}. On the other hand, in many application scenarios, wireless sensors can only be powered by batteries. If the battery of a wireless sensor is depleted, the data it is responsible for collecting may be missing, thereby affecting the integrity of target regional data. Therefore, an effective global solution that reduces energy consumption while prolonging the overall operation time in the IoT is indispensable. Fig. \ref{fig1} depicts an example of edge-assistant IoT networks. In typical scenarios, redundantly deployed sensors collect data in the target area. Subsequently, data is formatted and compressed on intelligent control devices with limited resources for simple inferential analysis and computation. Finally, base stations integrate the data into the edge servers to support upper-level applications such as data analytics.

The existing efforts, however, focus on either communication or computing, trying to optimize the energy consumption \textbf{from the upper level}.
To diminish communication energy consumption, data can be preprocessed and compressed, or the number of communications can be decreased, thus reducing the volume and number of transmissions~\cite{chen2021energy, xu2019energy, stojkoska2017data}.
To lower computing energy consumption, the complexity of the algorithm and architecture is optimized, or the location of computing loads and resource allocations are dynamically adjusted to reduce energy consumption~\cite{jiang2023energy, feng2021energy, zhao2017energy}.

Despite their advantages, neglecting the redundant sensing and storage of the underlying data source cannot fundamentally handle the ``butterfly effect". Fortunately, the low-power sensor has gained traction in the market, and it has a sleep mode to reduce the data reading range and unnecessary energy consumption. We note that the underlying scheme can be orthogonal with the above upper-level energy optimization techniques and act as the precursor step of the existing methodologies. 
Additionally, the heterogeneity of resources, workloads, and available energy among devices in the IoT may lead to unequal operational duration. When collecting data from an area, regional data incompleteness is prevalent due to the depletion of power in individual devices. Therefore, it is necessary to save the energy of devices on the verge of exhaustion to ensure the integrity of regional data.

To fulfill the aforementioned energy optimization requirements in IoT, we propose Senses, an online energy-saving solution for edge IoT networks. 
Senses reduces the energy consumption of edge IoT systems in \textbf{a bottom-up and fine-grained manner}. Specifically, Senses continuously monitors each control device's battery level, workload, and sensor coverage. It adopts a Multi-Agent Reinforcement Learning (MARL) scheme to reduce the data volume. At the sensor level, Senses deletes redundant data by dynamically adjusting sensor coverage. At the controller level, Senses partitions the data and eliminates redundant data. At the edge server level, Senses balances the workload of edge servers by employing strong and weak agent policies, integrating data with the desirable edge server. 
At the global level, considering the heterogeneity of control devices, Senses balances the operational duration of the control devices to prolong the operational duration of the whole IoT device network. During the adjustment of sensor coverage in Senses, robust control devices (with more available energy and lower workload) tend to share the sensor coverage of dying devices (with less available energy and higher workload). The major contributions of this paper can be summarized as follows:

\begin{itemize}

\item We propose Senses, a novel online energy saving solution for edge-IoT device networks, which jointly adapts the sensor, controller, edge server, and load distribution.
As a result, redundant data can be avoided at the data source level, saving energy in subsequent transmission and computation processes.

\item For general scenarios with heterogeneous computing resources, energy, and workloads in control devices, we further balance the operational duration of control devices and prolong the overall operational duration of edge-IoT device networks. 

\item Theoretical analysis and numerical results demonstrate that Senses achieves the wanted design rationales. It outperforms the baseline regarding energy consumption in data collection, transmission, and processing under the same operational duration. Moreover, Senses is equally outstanding in preventing severe regional data gaps.

\end{itemize}

The remainder of this paper is organized as follows. Section \ref{sec2} introduces the background and related work. Section \ref{sec3} presents the system model and the problem assumptions. Section \ref{sec4} proposes the Senses algorithm. Section \ref{sec5} reports the evaluation results. Finally, Section \ref{sec6} concludes the paper.

\section{Related work}\label{sec2}

\subsection{Energy Optimization in Edge Computing}

In the edge IoT, computing-induced energy consumption is mainly related to the device's system architecture and workload.
Tasks with high real-time demands may require faster processors and more computing resources, leading to heightened energy consumption~\cite{chen2020deep}. Moreover, the energy consumption for operating on different IoT devices can vary. Consequently, numerous studies demonstrate that the deployment method of edge servers and the location of computing task offloading should consider the energy consumption optimization ~\cite{lu2020edge,yang2020deep,bushehri2023deep}.
Li et al. \cite{li2018energy} study the energy-aware edge server placement problem and design a particle swarm optimization-based energy-aware edge server placement algorithm to find a more efficient low-energy placement scheme. Xu et al.~\cite{xu2017online} propose an efficient reinforcement learning-based resource management algorithm that addresses dynamic workload offloading and edge server provisioning for energy-harvesting MEC systems. 
Li et al. \cite{li2021energy} focus on the problem of minimizing energy consumption in mobile edge computing systems, where both local and edge computing capabilities are limited and may lead to task discarding due to delay conflicts in both binary offload and partial offload modes. Yu et al. \cite{yu2021rsu} focus on the critical issues of joint task offloading and energy scheduling in green multilayer edge computing systems. The task execution cost is minimized by jointly considering the system cost, which includes latency, energy consumption, and cloud rental cost.

Optimizing the architecture is another approach, and a plethora of green architectures have been explored to reduce energy consumption during the IoT operating process~\cite{moghaddam2018fog,nakhkash2019analysis}. By considering the unique characteristics of IoT devices and their applications, architects can design energy-saving architectures that meet the requirements of different scenarios. 
Ammad et al. \cite{ammad2020novel} propose a novel architecture that considers the heterogeneity of the main sources of energy consumption, server collaboration, and traffic scheduling strategies. Significant energy savings can be achieved by intelligently distributing tasks and optimizing resource allocation.
Xiang et al. \cite{xiang2021energy} also emphasize the importance of architecture optimization for energy efficiency in the Internet of Things. They present an architecture combining edge and cloud computing to minimize energy consumption. 
Wang et al. \cite{wang2020hierarchical} discuss a hierarchical architecture for IoT applications that aims to optimize energy consumption. By dividing the network into different layers and assigning tasks based on energy requirements, this architecture can significantly reduce energy consumption while maintaining performance.

\subsection{Energy Optimization in Communications}
The communication-induced energy consumption in IoT is chiefly related to the volume of data~\cite{has2024efficient,zhang2022learning}. Thus, effective data compression is crucial for reducing the power consumption of IoT devices.
Data compression primarily centers on enhancing accuracy, reducing latency, minimizing distortion levels, and modifying more robust data compression technologies~\cite{al2021energy,chen2023deep}. The mainstream data compression schemes fall into two categories: lossy and lossless.
Lossy compression typically achieves a high compression ratio, significantly reducing the data volume, lowering transmission costs, and enhancing transmission efficiency~\cite{deepu2016hybrid,wang2016tree}. However, it inevitably leads to a certain degree of irreversible information loss~\cite{abdellatif2018user,chen2019fog}. 
To this end, lossless compression is further put forward, guaranteeing data integrity during compression and decompression without any information loss~\cite{huffman1952method,ziv1978compression,liang2014efficient}. Some lossless data compression schemes are dedicated to IoT networks. Nevertheless, compared to lossy compression algorithms, it typically has a lower compression ratio. Moreover, the algorithm is more complex and demands more computation slots~\cite{kang2017energy}.
In addition, there are also some model compression methods applied in scenarios involving transmission models (such as distributed machine learning, federated learning, Large Language Models (LLM), etc.)~\cite{zawish2022energy,cheng2017survey,shah2021model,ma2023llm}, reducing the amount of transmitted data by pruning the model.

Existing works mainly try to optimize the energy consumption from the upper level by tuning the service offload on each control device or compressing the data to transmit. In contrast, our work considers optimizing IoT device network energy consumption under edge-IoT federation from a different perspective, recommending a bottom-up and fine-grained optimization by jointly considering sensors, control devices, and edge servers.

\section{Problem Formulation}\label{sec3}
This section first states the problem with fundamental assumptions in Section \ref{formu-1}. After that, we analyze the energy consumption generated by the data collection, transmission, and processing. Finally, we describe the optimization objective in Section \ref{formu-4}. For clarity, the major notations used in this paper are listed in Table \ref{tab:table1}.

\begin{table}[!t]
    \caption{Summary of notations\label{tab:table1}}
    \centering

    \begin{tabular}{cc}
    \toprule
    \emph{Notation} & \emph{Description} \\
    \midrule
     $K, M$&number of sensors and control devices,\\
     &respectively\\
      $S$&the set of sensor terminals \\
      $C$&the set of control devices \\
      $D$&the data set from the control device obtained from\\ &the sensors \\
      $\xi$&the set of edge servers $E_j$\\
      $ES_j$&storage resources of the edge server $E_j$\\
      $EC_j$&computing resources of the edge server $E_j$\\
      $\phi$&The set of service requests\\

      \hline
      $R_i(\tau)$&the coverage of the sensor $s_i$ at time $\tau$\\
      $SW_i$&the relationship between the operating state\\
      &and the operating current of each sensor\\
      $U_i$& the working voltage of sensor $s_i$\\
      $h_\phi$&transfer vector between control devices and sensors\\
      $h_{i}^{c}$&transition matrix between edge server and\\ &control devices\\
      $h_i^{e}$&transition matrix between edge servers\\
      $\iota(\tau$)&the state of the battery\\
      $W^n(\tau), W^j(\tau)$&the workload of each control device and edge server\\
      \hline
      $o(\tau)$&environment state of tasks in the edge server at time $\tau$ \\
      &set from $a$ to $b$\\
      $a(\tau)$&action taken by the agent at time $\tau$\\
      $r(\tau)$&reward taken by the agent at time $\tau$ \\
      $A(\tau)$&advantage function\\
    
      $\theta_1, \theta_2$& parameters of the actor net and critic net, respectively\\
      \bottomrule
    \end{tabular}
\end{table}

\subsection{Problem Statement}\label{formu-1}
We consider the Edge IoT networks within the appropriate geographical scope. Then assume that there are $K$ sensors in this network, denoted by $S=\{s_1,s_2,...,s_K\}$, $s_i(1 \le i \le K)$ represents the $i^{th}$ sensor. Sensors can transmit data to the control device through different communication protocols, i.e., RS$232$ and AD\--DA. The control device can shield the communication protocol differences to integrate and generalize data. In our scenarios, the control device can be represented as $C=\{c_1,c_2,...,c_M\}$, $c_n(1 \le n \le M)$ represents the $n^{th}$ control device. Time is evenly discretized into brief periods. In each cycle $\tau$, a control device collects the data information of sensors to form a dataset $D=\{d_1,d_2,...,d_K\}$. Furthermore, control devices usually also undertake tasks such as motor and servo driving. For the convenience of calculation, the energy consumption of these drivers is not considered. Due to fewer computing and storage resources, the control device generally transmits sensor data to the edge server for high-precision processing. We define these edge servers as $\xi=\{E_1,E_2,...,E_N\}$, $E_j(1 \le j \le N)$ represents the $j^{th}$ edge server. 

Compared to the cloud, edge servers can provide services with lower latency since they are placed closer to users. However, the computing and storage capacity of each edge server is limited. If we want to transfer data to the edge server for high-precision processing, the resource capacity of each edge server needs to be recognized. We define the computing and storage resource capacity on each edge server $E_j$ as $EC_j$ and $ES_j$. In addition, given the high-quality communication brought by the development of information technology, we assume that control devices could maintain a stable connection with the same edge server during data processing and transmission.

\subsection{Energy Consumption}\label{formu-2}
Sensors consume energy during data collection, typically supplied by the battery carried by control devices. The energy consumption of a sensor is mainly influenced by its current working state (data collection range). We define the coverage of the sensor at time $\tau$ as:
\begin{equation}
R(\tau)=[R_1(\tau),R_2(\tau),...,R_K(\tau)]
\end{equation}
$R_i(\tau)$ is a rational number, denotes the working coverage of the sensor $s_i$, and the coverage constraint is:
\begin{equation}
0 \leq R_i(\tau) \leq R_{Max}(\tau)
\end{equation}
where $R_{Max}(\tau)$ indicates maximum sensor coverage. Since most sensors are powered by direct-current power, the voltage is stable, and the operating state affects the current. The relationship between the operating state and current of each sensor is expressed as:

\begin{equation}
SW_i=[SW_1,SW_2,...,SW_K]
\end{equation}
where $SW_K$ denotes the known nonlinear formula of the sensor $s_K$ at the factory. Then, we define the energy consumption of each sensor in the time slot $\tau$ as:
\begin{equation}
e_i^s(\tau)=SW_i(R_i(\tau))U_i\tau
\end{equation}
$U_i$ is the operating voltage of sensor $s_i$, usually a fixed constant. $e_i^s(\tau)$ denotes the energy consumption caused by sensor $s_i$ collecting data within time $\tau$. After data collection, sensors should transmit data to control devices through specific communication protocols. To surface the relationship between data transmission of control devices and sensors, we define:
\begin{equation}
h_\phi=[h_i^1,h_i^2,...,h_i^M]
\end{equation}
$h_i^M$ is a binary value, with $h_i^M=1$ denoting that sensor $s_i$ transmits data to the control device $c_M$. $h_i^M=0$, otherwise. Then, we define:
\begin{equation}
B_\phi=[B_i^1,B_i^2,...,B_i^M]^T
\end{equation}

\begin{equation}
P_\phi=[P_i^1,P_i^2,...,P_i^M]^T
\end{equation}
where $P_i^M$ denotes the data transmission power that sensor $s_i$ transmits data to control device $c_M$. $B_i^M$ denotes the data transmission rate at which sensor $s_i$ transmits data to control device $c_M$. $B_\phi$ and $P_\phi$ are mainly affected by communication protocols. Then we define:

\begin{equation}
e_{i,n}^{tran1}=\frac{|d_i|}{B_\phi \circ h_\phi} (P_\phi \circ h_\phi)
\end{equation}
the symbol $\circ$ represents the operation of the inner product, $|d_i|$ indicates the size of the sensor data $d_i$, which is mainly affected by the collection range of the sensor. $e_{i,n}^{tran1}$ denotes the energy consumption of data collected by sensor $s_i$ within time $\tau$ and transmitted to control device $c_n$. After receiving the sensor data, control devices may conduct general data processing due to communication protocol heterogeneity of data formats. Therefore, we define this part of processing energy consumption as:
\begin{equation}
e_{i,n}^{comp}=k|d_i|(f_i^{comp})^2
\end{equation}
where $f_i^{comp}$ denotes the local computing power of the control device to process the sensor data $s_i$. $k$ is an energy factor. After data processing, the control device should transmit data to the edge server to support relevant applications. At this stage, the control device may also transmit other data, such as motor speed and temperature, which are affected by more factors. Hence, we do not consider optimizing the calculation and transmission of these data here. Then, given that some duplicate data will not be transmitted to the edge server, we define the following:
\begin{equation}
h_i^c={
\left[ \begin{array}{cccc}
h_i^{11} & h_i^{12} & \cdots & h_i^{1N}\\
h_i^{21} & h_i^{22} & \cdots & h_i^{2N}\\
\cdots & \cdots & \cdots & \cdots\\
h_i^{M1} & h_i^{M2} & \cdots & h_i^{MN}
\end{array} 
\right ]}
\end{equation}
$h_i^{MN}$ is a binary value, with $h_i^{MN} = 1$ denoting that the control device $c_M$ will transmit data $d_i$ to the edge server $E_N$, $h_i^{MN} = 0$, otherwise. Then, we define:
\begin{equation}
B_\theta(\tau)={
\left[ \begin{array}{cccc}
B^{11}(\tau) & B^{12}(\tau) & \cdots & B^{1N}(\tau)\\
B^{21}(\tau) & B^{22}(\tau) & \cdots & B^{2N}(\tau)\\
\cdots & \cdots & \cdots & \cdots\\
B^{M1}(\tau) & B^{M2}(\tau) & \cdots & B^{MN}(\tau)
\end{array} 
\right ]}
\end{equation}
where $B^{MN}(\tau)$ denotes the data transmission rate at which control device $c_M$ will transmit data to edge server $E_N$ at time $\tau$. Next, we define:
\begin{equation}
P_\theta(\tau)={
\left[ \begin{array}{cccc}
P^{11}(\tau) & P^{12}(\tau) & \cdots & P^{1N}(\tau)\\
P^{21}(\tau) & P^{22}(\tau) & \cdots & P^{2N}(\tau)\\
\cdots & \cdots & \cdots & \cdots\\
P^{M1}(\tau) & P^{M2}(\tau) & \cdots & P^{MN}(\tau)
\end{array} 
\right ]}
\end{equation}
$P^{MN}(\tau)$ indicates the data transmission power that data be transmitted from control device $c_M$ to edge server $E_N$ at time $\tau$. We express the energy consumption of sensor data transmission from the control device to the connecting edge server as:
\begin{equation}
e_{i,n,j}^{tran2}=\frac{|d_i^c|}{ || B_\theta(\tau) \bigotimes h_i^c||_1} (|| P_\theta(\tau) \bigotimes h_i^c||_1) \tau
\end{equation}
where the symbol of $\bigotimes$ denotes the Hadamard product, which is used to multiply corresponding elements in two matrices. $| | \bullet | |_1$ means the L$1$-norm of matrix. $|| B_\theta(\tau) \bigotimes h_i^c||_1$ denotes the data transmission rate of the link from $c_n$ to $E_j$, and $|| P_\theta(\tau) \bigotimes h_i^c||_1$ denotes the data transmission power of the link from $c_n$ to $E_j$.

In some application scenarios, all data in a region need to be centralized to a computing node for reasoning and analysis. We observe that each edge server may only be connected to a portion of control devices and define the energy consumption of data transmission between edge servers as follows:
\begin{equation}
e_{i,n,j}^{tran3}=\frac{|d_i^c|}{||B_i^e(\tau) \bigotimes H_i^e||_1} (||P_i^e(\tau) \bigotimes h_i^e||_1)
\end{equation}

\subsection{Other Factors}\label{formu-3}
At an arbitrary time $\tau$, the state of the battery on control devices can be modeled as:
\begin{equation}
\iota (\tau)= <SoE(\tau),SoC(\tau),DoD(\tau)>
\end{equation}
where the notation $SoE$ represents the state of effective capacity of the battery, which is expressed as a percentage of the initial capacity. $SoC$ means the state of charge of the battery, which is expressed as a percentage of the current effective capacity and indicates the current energy stored. $DoD$ is the depth of discharge of the battery, which is expressed as a percentage of the current effective capacity and indicates how much power the battery releases. We define the battery set in the Edge IoT networks as $L={\iota_1(\tau),\iota_2(\tau),...,\iota_Q(\tau)}$. Besides, to prevent each battery from over-discharging, we define the upper and lower bounds for $Soc$:
\begin{equation}
Soc_{min} \leq Soc(\tau) \leq Soc_{max}
\end{equation}

The rechargeable scenarios like solar batteries are not elaborated in detail here, and the power increase is expressed in $SoC(\tau)$. Additionally, control devices typically have workloads of varying degrees (i.e., determined by the number of connected sensors and the size of the driving motor). Likewise, edge servers, similar to control devices, inevitably have different workloads. We set bounds for the workload of each control device $W^n(\tau)$ and edge server $W^j(\tau)$ as follows:

\begin{equation}
W^n_{min} \leq W^n(\tau) \leq W^n_{max}
\end{equation}

\begin{equation}
W^j_{min} \leq W^j(\tau) \leq W^j_{max}
\end{equation}

\subsection{Optimization Objective: Minimizing the Total Energy Consumption}\label{formu-4}
In our scenario, as described above, the entire process of sensor data transmission can be expressed as follows. For an arbitrary time slot $\tau$, sensors consume $e_i^s(\tau)$ energy to collect data and then send the data to the connected control device with energy consumption $e_{i,n}^{tran1}$. Subsequently, the data is generalized on the control device through energy consumption $e_{i,n}^{comp}$ to form a data set, and the data will be determined on the control device whether to go to the edge server. After that, the control device selects an appropriate edge server to transmit the data through energy consumption $e_{i,n,j}^{tran2}$. Furthermore, some comprehensive data are centralized to one computing node for processing through energy consumption $e_{i,n,j}^{tran3}$. The relevant computations to support the application are performed after the data arrives at the destination edge server. This part is not discussed in this paper.

With limited energy, competition for processing and transmitting sensor data is inevitable. If our goal is to optimize the energy consumption of data transmission and processing of each sensor, there will be many contradictions. Therefore, we minimize the average energy consumption of all sensor data transmission and processing with the following optimization objectives:

\begin{equation}
\label{eq19}
\begin{aligned}
&\min\sum^M_{ n=1}\sum^K_{i=1}(e_{i,n}^{tran1}+e_{i,n,j}^{tran2}+e_{i,n,j}^{tran3}+e_{i,n}^{comp}+e_{i}^{s}(\tau))\\
&\qquad s.t.\quad (16) \sim (18)
\end{aligned}
\end{equation}

In the above optimization problem, the objective function is to minimize the total energy consumption of the IoT device networks, which consists of the energy consumption generated by transmission, computation, and data collection. The constraints set limits on the resources and battery power provided by the edge servers, as well as IoT control devices. $d_i$ and $d_i^c$ are the unknown variables to be solved. However, we face several challenges in solving the above optimization problem. First, the number of IoT control devices and edge servers in the real world has increased from hundreds to thousands. Traditional optimization algorithms take a long time to develop the best strategy for sensor coverage. This challenge leads to a huge search space when solving optimization problems, and therefore, force-based search methods are prohibited. Furthermore, offline optimization solutions may be far from reality due to dynamic battery power consumption and the nature of network resources. Online solutions for real-time estimation of the network conditions should be considered.

To tackle the above challenges, we propose an online scheme, Senses, based on MARL techniques in the following section.

\begin{algorithm}[t]
\caption{Senses}
\label{alg1}
\small
\begin{algorithmic}[1]
\STATE \textbf{Processing at the Edge Server}
\FOR {each updates}
\WHILE {the servers have not collected all the control device data}
\STATE Waiting for the control device to transfer data
\ENDWHILE
\STATE Centralise and classify data
\STATE Distribute the updated coverage to all control devices
\ENDFOR
\STATE \textbf{Processing at the control device $c_i$}
\STATE Receive the sensor coverage from the edge server
\FOR {each updates}
\STATE Adjust the coverage of sensor data collection
\STATE Receive sensor data
\STATE Partition the data and eliminate redundant data
\STATE Decode and format conversion

\STATE Push battery level $Soc(\tau)$, current load $CL$, future load $FL$, and GPS data $G$ to servers
\ENDFOR

\end{algorithmic}
\end{algorithm}

\begin{figure}%
\centering
\includegraphics[scale=0.9]{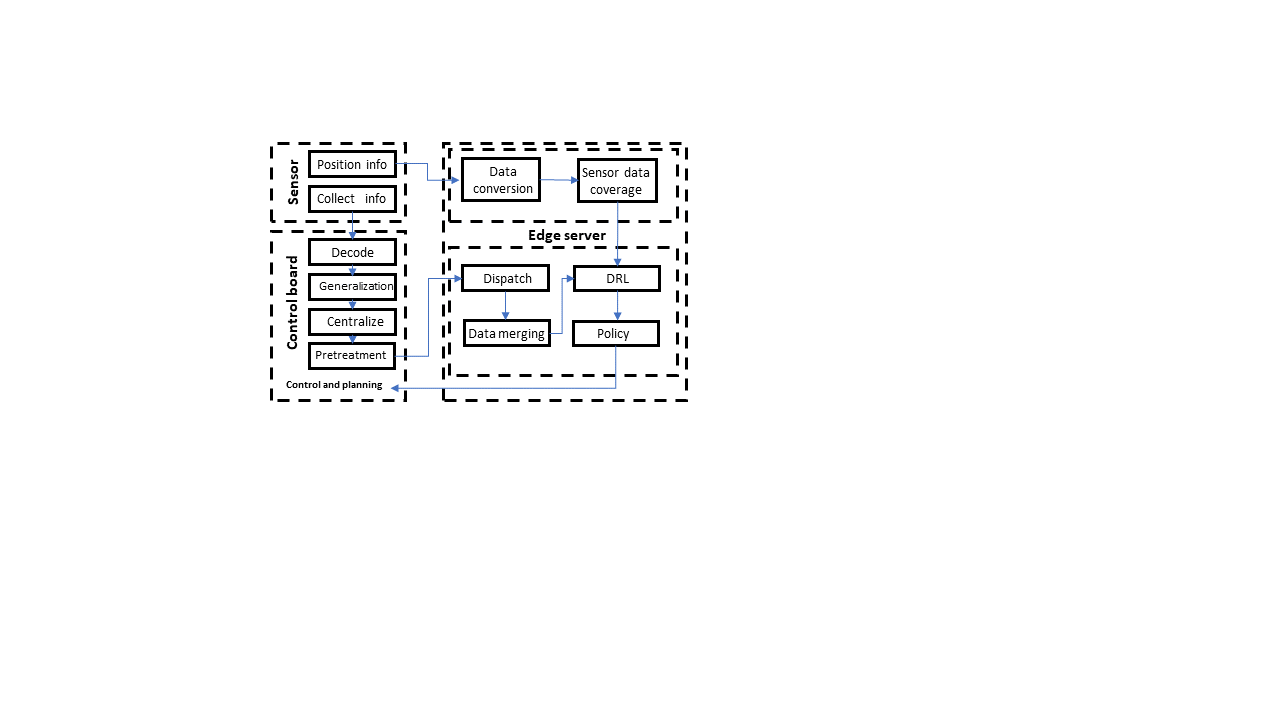}
\caption{Edge IoT network workflow.}%

\label{fig2}%
\vspace{-0.2in}
\end{figure}

\section{The Senses Solution based on Multi-Agent Reinforcement Learning} \label{sec4}
The breakthroughs in Multi-Agent Reinforcement Learning (MARL) in recent years have provided an effective empirical approach~\cite{zhang2019proximal,liu2024computation}. MARL enables multiple agents to interact and learn within a dynamic environment. By leveraging historical experience, it makes better decisions by adapting to the current state of the environment. We contend that MARL could be an excellent solution for optimizing the energy consumption of networks in edge IoT, given its capability to handle time variability (e.g., dynamic IoT network states in uncertain environments). Subsequently, we elaborate on the MARL-based optimization algorithm in detail.

Fig. \ref{fig2} depicts the data collection workflow by the sensor, data decoding and format conversion by the control device, and decision-making by the edge server. Under this workflow, we describe the MARL-based optimization algorithm in Algorithm. \ref{alg1}. For each update, the edge server first accepts and classifies the IoT control device data (Line 2-6 in Algorithm. \ref{alg1}). The sensors' data are used for upper-layer applications. The operational information of control devices, sensors, and edge servers are used for policy-making in the MARL algorithm described in detail in Algorithm \ref{alg2}.
Finally, the edge server transmits the updated policy to the control device (Line 7 in Algorithm. \ref{alg1}).
The control device uses measurement tools to determine the current battery level, GPS data, current load, and future load. When the control device receives the sensor data, it first transcodes the data into a specific format for the individual sensor communication and adjusts the sensor collection data range. (Line 10-15 in Algorithm. \ref{alg1}).
Finally, control devices transmit sensor data and operational information to the servers (Line 16 in Algorithm. \ref{alg1}).

\begin{figure}[t]%
\centering
\includegraphics[scale=0.45]{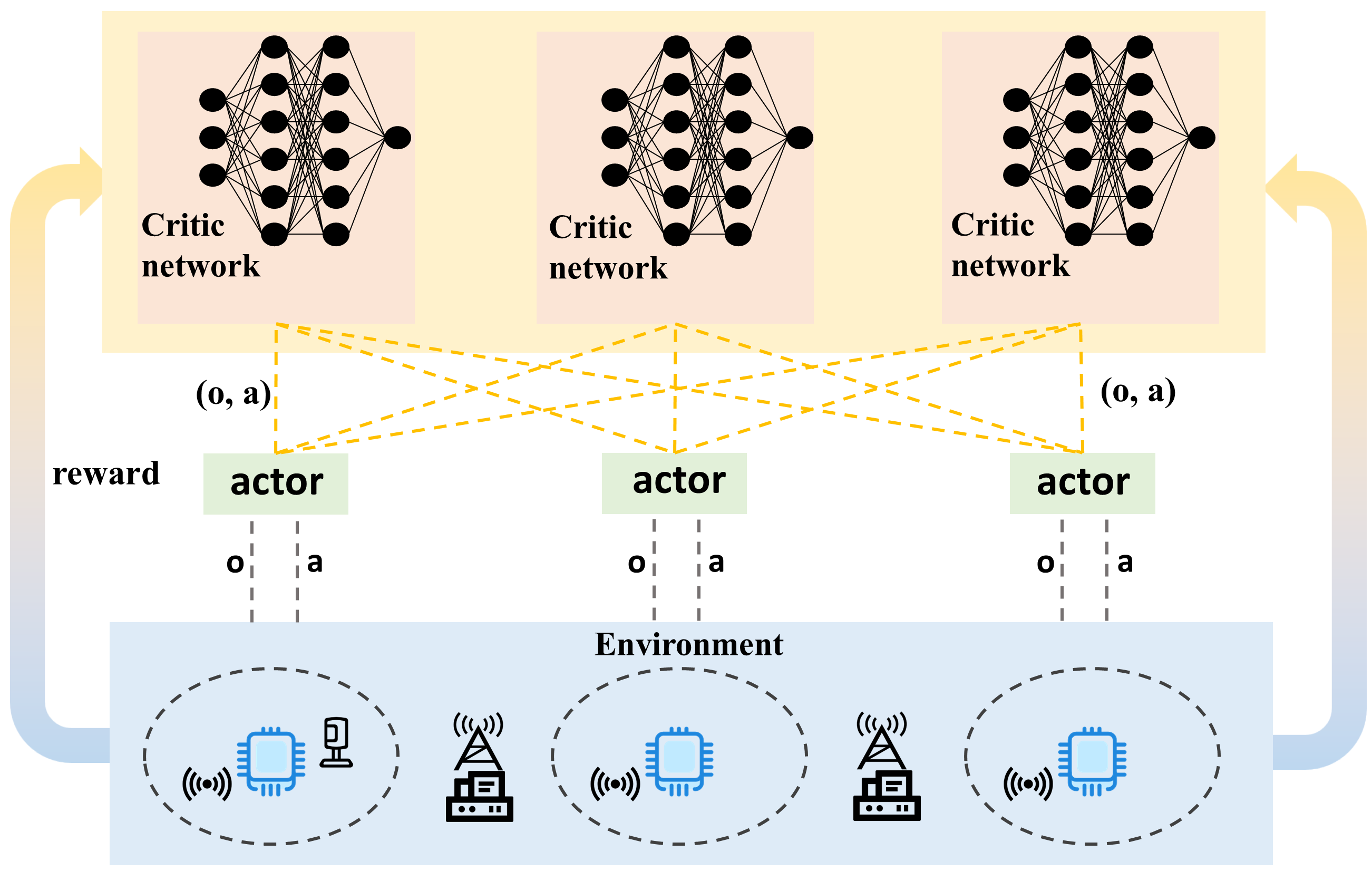}
\caption{The framework of MARL-based solution at the edge.}%
\label{fig3}%
\vspace{-0.2in}
\end{figure}

Multi-agent Proximal Policy Optimization (MAPPO) is a variant of the PPO algorithm applied to multi-agent tasks, adopts an actor-critic architecture, and can effectively handle the interaction and collaboration between agents\cite{yu2022surprising}. As an on$\--$policy algorithm, MAPPO has significantly high algorithm operation efficiency and comparable data sample efficiency. It is more adaptable to edge conditions with limited computing resources.
Based on the MAPPO algorithm, we propose Senses, a multi-agent reinforcement learning method with strong and weak agents. Specifically, Senses captures the load of the edge server in real-time. When the load is large, the critic network encourages agents to become weak agents and reduces the desire to participate in sensor range adjustment. When the load is small, the opposite is true. In addition, Senses adopts centralized training and distributed execution, allowing agents to share some global information during the training process to learn collaborative strategies better. Agents can maintain independence during the execution process, improving the scalability and adaptability of the algorithm.
Fig. \ref{fig3} depicts the framework and main components of the MARL-based approach, and the design of the components is explained in detail as follows.

\textbf{Agent}: Agent is placed on the edge server. Each agent is responsible for observing the environment, choosing actions according to its policy, and receiving rewards. Specifically, at the beginning of each time slot $\tau$, it determines the sensor data coverage policy according to the current state $o(\tau)$ of the environment. During training, the agent will iteratively loop the above process to improve decision-making. The goal of the agent in this paper is to find an optimal sensor coverage control strategy through interaction with the environment and other agents, so as to minimize the total energy consumption of the entire cycle while prolonging the network operational duration.

\begin{algorithm}[t]
\caption{ Adjustment algorithm for sensor coverage with MARL}
\label{alg2}
\small
\begin{algorithmic}[1]
\REQUIRE Initialize the parameters of actor network $\theta_{1}$ and the parameters of critic network $\theta_{2}$.
\ENSURE Optimal sensor coverage policy 
\STATE Initialize learning rate $\alpha$, parameter $\lambda$, GAE discount factor $\gamma$, clip range $\epsilon$
\WHILE {step $\leq$ stepmax} 
\STATE Set data buffer
\FOR {$\tau=1$ to $T$ }
\FOR {all agents}
\setlength{\abovedisplayskip}{0pt}
\setlength{\belowdisplayskip}{0pt}

\STATE actor generates action $u_{\tau}^{a}$ for state $o_{\tau}^{a}$ based on policy $\pi_{\theta_{1}}(u_{\tau}^{a}\mid o_{\tau}^{a})$

\ENDFOR
\STATE Execute actions $u_{\tau}$, observe $r_{\tau}$, $o_{\tau+1}$
\STATE Replay buffer collecting $o_{\tau},u_{\tau},r_{\tau},o_{\tau+1}$
\STATE policy $\pi_{old}=\pi$

\STATE Compute advantage estimate $A_{\tau}$ via GAE

\IF {buffer \textgreater batchsize}
\STATE Update $\theta_{1}$ of actor network
\STATE Update $\theta_{2}$ of critic network
\ENDIF

\ENDFOR
\ENDWHILE
\end{algorithmic}
\end{algorithm}

\textbf{Advantage function}: Advantage function $A(\tau)$ measures the degree of advantage of perform a action $a(\tau)$ in a state $o(\tau)$ compared to performing the average action. $A(\tau)$ is positive, indicating that action $a(\tau)$ is better than the average, and its selection probability should be increased. $A(\tau)$ is negative, which means action $a(\tau)$ is relatively worse, and its selection probability should be reduced.

We employ Generalized Advantage Estimation (GAE) to calculate the advantage function, which combines the $\lambda$-return algorithm~\cite{daley2019reconciling} to extend state estimation to $n$ steps, thereby balancing the variance and bias in reinforcement learning.
The advantage function for the agent can be expressed as:
\begin{equation}
\label{eq20}
    A_(\tau)=\sum_{l=0}^{\infty}(\gamma \lambda)^{l}\delta_{\tau+l}
\end{equation}
where $\gamma$ is a discount factor, $\delta_{\tau}$ indicates the advantage estimation in time slot $\tau$.
\begin{equation}
    \delta_(\tau)=r_{\tau}+\gamma V(o_{\tau+1})-V(o_{\tau})
\end{equation}
where $V(o_{\tau})$ is a value function for the critic network to estimate the expected reward in the state $o_{\tau}$. $r_{\tau}$ is the reward in time slot $\tau$.
According to eq. \ref{eq19}, the reward is formulated as follows:
\begin{equation}
    r_(\tau)= \mathbb{E}(e^{cost}+T_{max}-L)
\end{equation}
where $\mathbb{E}$ is the value function, $e^{cost}$ means the energy consumption, $T_{max}$ is the maximum operational duration, and $L$ is the workload of edge server. Therefore, the return can be expressed as:
\begin{equation}
    G(\tau)=\sum_{l=0}^{\infty}\gamma ^{l}r_{\tau+l+1}+A_{\tau}
\end{equation}

\textbf{Critic network}: Critic network is responsible for assessing the quality of the actions taken by agents. It estimates the value function, which represents the expected cumulative reward that an agent can obtain starting from a given state. The critic network provides a benchmark for the evaluation of policies. By comparing value estimations in different states, agents can determine the quality of the current policy and decide whether to adjust the policy to obtain higher rewards. In this paper, our critic perceives overall environmental changes, including the working information of sensors, control devices, and edge servers within the area.

\begin{equation}
\label{eq24}
    L(\theta_{2})=-\sum_{\tau=0}^{T}(G(\tau)-V_{\pi_{\theta_{2}}(o_{\tau})})^{2}
\end{equation}

\textbf{Actor network}: Actor network is responsible for generating the action policy of the agent. Based on the observation from the environment, it outputs the probability distribution of performing various actions in the current state, and the agent selects a specific action by sampling from this probability distribution. The loss function of the actor network is typically constructed based on the policy gradient:

\begin{equation}
\label{eq25}
    L(\theta_{1})=\frac{1}{N}\sum_{i=1}^{N}min(r_{\theta_{1},i}A_{i},clip(r_{\theta_{1},i},1-\epsilon,1+\epsilon)A_{i})
\end{equation}
where $r_{\theta_{1},i}=\frac{\pi_{\theta_{1}}(a_{i} \mid o_{i})}{\pi_{\theta_1old}(a_{i} \mid o_{i}))}$, $N$ is the number of agents, $\epsilon$ is a clip range parameter, $\epsilon \in [0,1]$.

This loss function updates the policy to improve the advantage function while maintaining a certain degree of proximity between the new and old policies, thereby enhancing the agent's performance.

\begin{figure}[t]%
\centering
\includegraphics[scale=0.35]{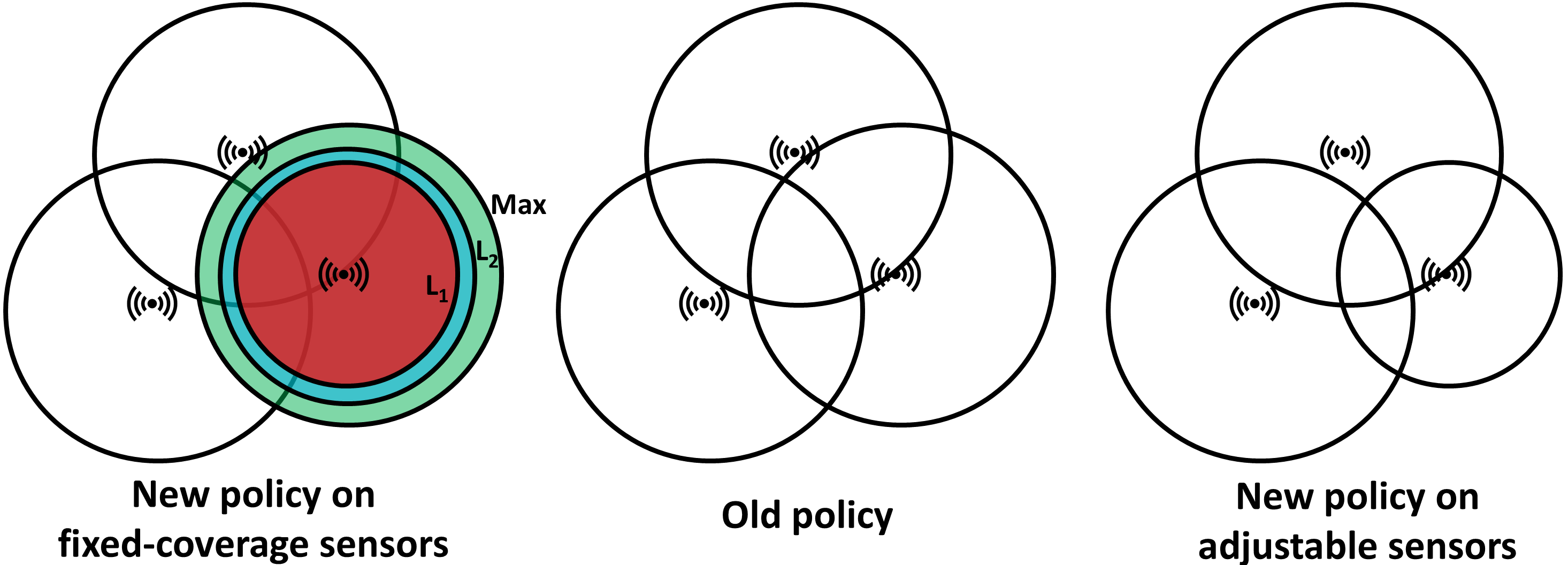}
\caption{The control strategy on fixed-coverage sensors (Left) and adjustable sensors (Right).}%
\label{fig4}%
\vspace{-0.2in}
\end{figure}

\textbf{Environment}: In this paper, the environment is constituted by entities of edge servers, control devices, and sensors. Agents can operate the environment state and perform sensor range adjustment actions based on the workload of edge servers and the operational information of control devices.

\textbf{Control strategy}: The coverage adjustment policy of sensors is output by the MARL on edge servers and executed by control devices. However, not all sensors have adjustable ranges. Therefore, for sensors with fixed operating coverage, we perform multi-layer zoning on the data after receiving sensor data through control devices, uploading the data in batches according to priorities. Specifically, the data of sensors with fixed operating coverage is divided into three priorities. Fig. \ref{fig4} visualizes an example of such a partition on a region of three sensors. The red area represents the highest priority, which includes the scope for which a sensor should have been adjusted. The blue area represents the second priority between the boundary $L_2$ and the maximum operating coverage. In practical situations, $L_2$ could be adjusted as a hyperparameter. The boundary $L_2$ in here is calculated by $\frac{max+L_1}{2}$. The green area represents the lowest priority. During the transmission process, the control device will send data to the edge server step by step according to priorities. When the edge server detects receiving all data in the area, the control device will stop forwarding the remaining sensor data to reduce the ``butterfly effect" caused by redundant data.

The adjustment algorithm for the Edge IoT networks is summarized in the Algorithm. \ref{alg2}. It first initializes the learning rate $\alpha$, parameter $\lambda$, GAE discount factor $\gamma$, clip range $\epsilon$ (Line 1 in Algorithm. \ref{alg2}). 
For all agents, based on the current state $o_{\tau}^{a}$, the actor network uses the policy network $\pi$ and parameter $\theta$ to calculate the action probability distribution $p_{\tau}^{a}$. Then, sample an action $u_{\tau}^{a}$ from the action probability distribution $p_{\tau}^{a}$ (Line 4-7 in Algorithm. \ref{alg2}).
After that, agent perform $u_{\tau}$, observe $r_{\tau},o_{\tau+1}$, and store the transition $o_{\tau},u_{\tau},r_{\tau},o_{\tau+1}$ in replay buffer (Line 8-10 in Algorithm. \ref{alg2}).
The agent calculates an advantage estimate via the GAE detailed in eq. \ref{eq20}.
With enough experience in the buffer, agents update the actor and critic networks based on eq. \ref{eq24} and eq. \ref{eq25}, and then repeat the above actions until the iteration stops (Line 11-14 in Algorithm. \ref{alg2}).

\section{Performance Evaluation} \label{sec5}
In this section, we first introduce the experiment setting in Section \ref{sec5:1}. After that, we evaluate the performance of Senses on the simulation and testbed in Section \ref{sec5:2} and \ref{sec5:3}, respectively.

\subsection{Experimental Settings}
\label{sec5:1}
\subsubsection{Simulation} In this experiment, we use a laptop as the simulation experiment environment (operating system: $64\--$bit Ubuntu $23.10$; RAM: $32$GB DDR$5$ $5200$MHz; CPU: Intel $i9\--13900H$ $2.6$GHz; GPU: NVIDIA GEFORCE RTX $4060$). We assume that there are $5$ edge servers, $28$ control devices, and $30$ sensors ($24$ adjustable and $6$ fixed coverage) in the edge IoT. Edge servers can communicate with each other. Control devices are randomly connected to edge servers via wired/wireless connections, support the communication protocol of sensors, can read and control sensors, and supply power to them. We assume in a $100\times100$ area, control devices are distributed based on a greedy method to ensure coverage of the entire area. Each edge server is connected to at least one control device. The range of sensors is regarded as a circle, and the sensing radius is between $15$ and $25$. The power on the control device is between $2000$ and $4000$ watts. Combined with the actual situation, we believe each edge server and control device may also have other tasks. We refer to the energy consumption formula of commercial sensors and ensure that the energy consumption of transmission and computation is positively correlated with the amount of data.

\begin{figure}[t]%
\centering
\includegraphics[scale=0.4]{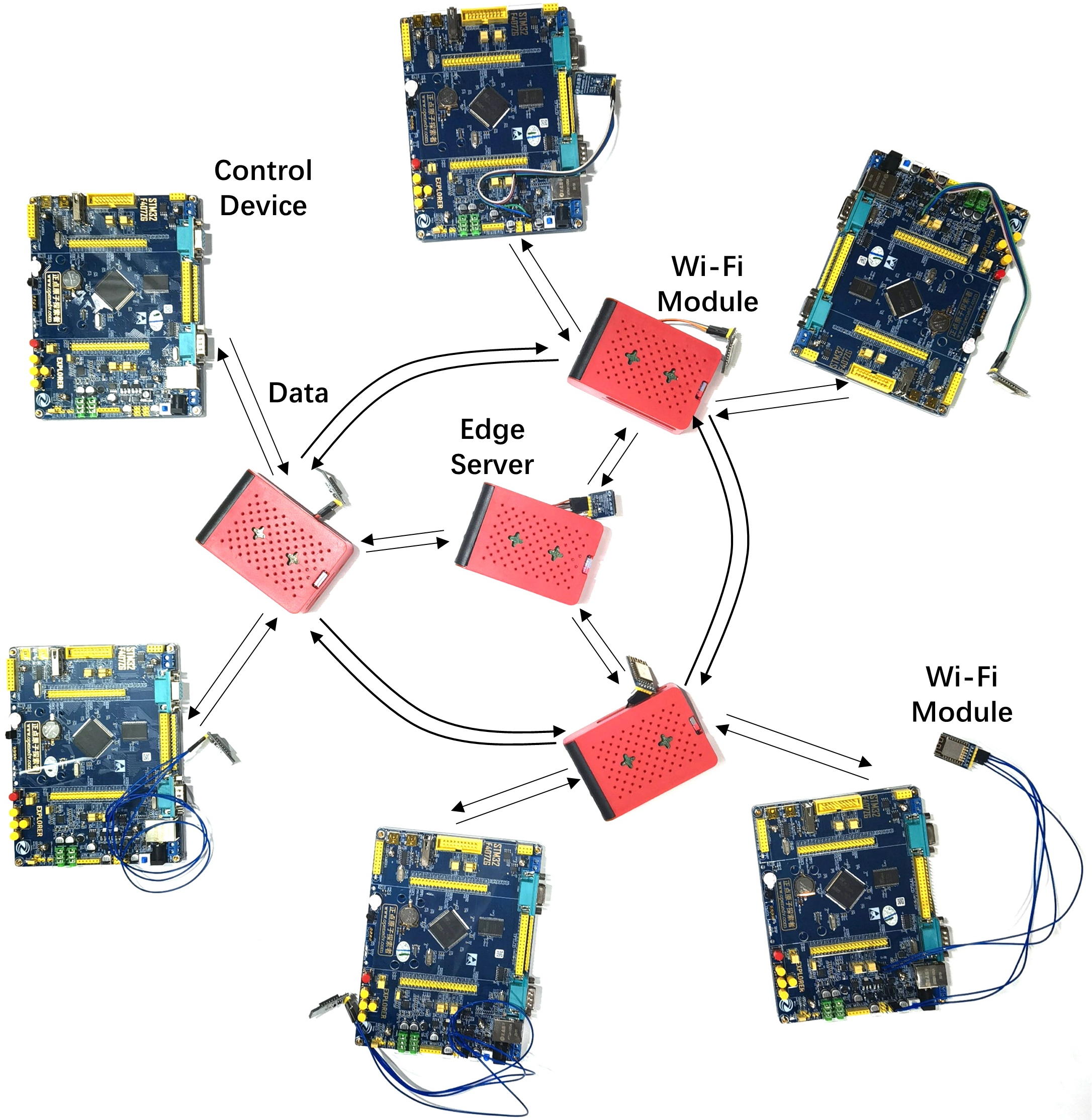}
\caption{The testbed platform with $4$ Raspberry Pi devices and $6$ STM32F407ZG microcontrollers.}%
\label{fig5}%

\end{figure}

\begin{figure}[t]%
\centering
\includegraphics[scale=0.35]{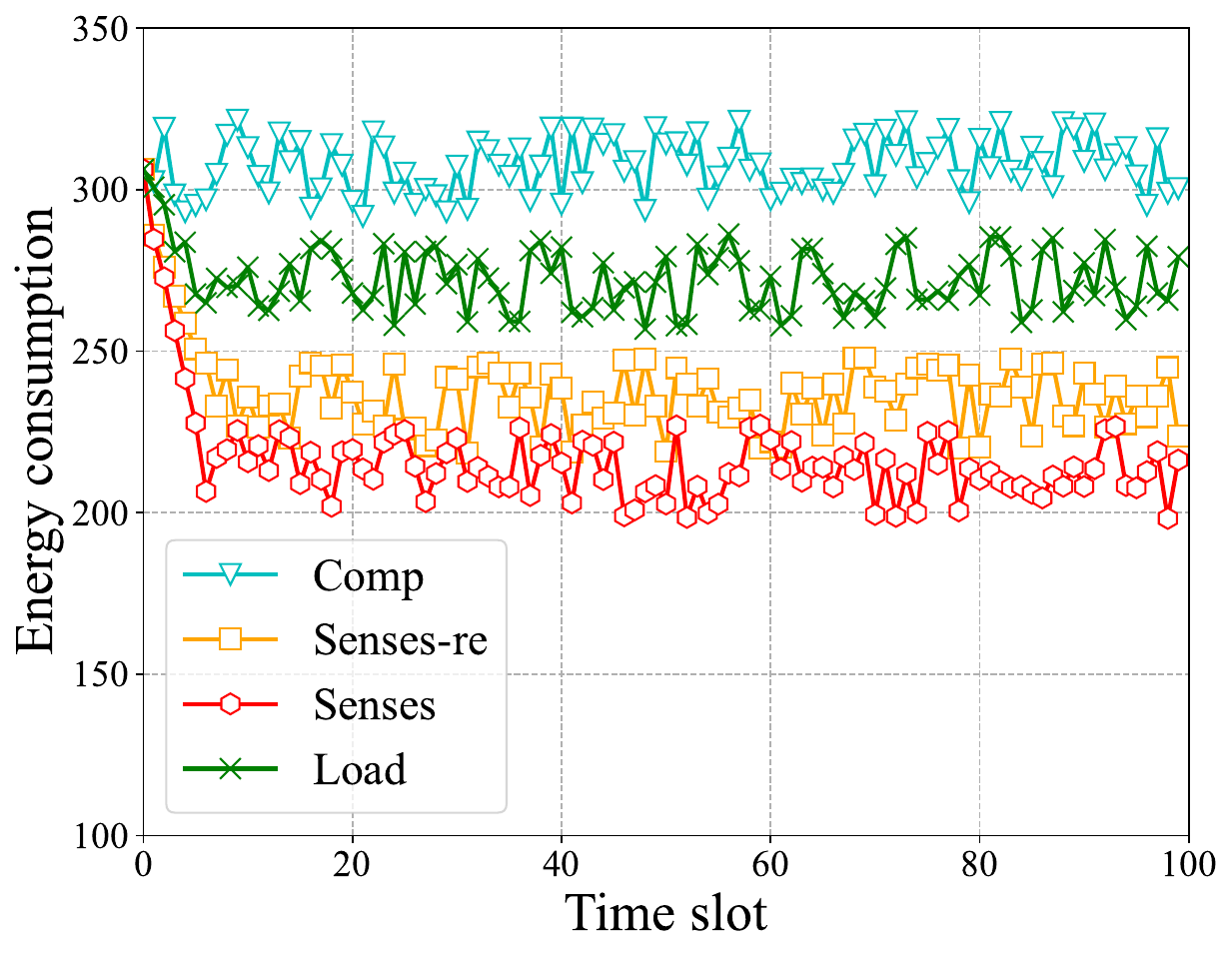}
\caption{Energy consumption of all control devices over the time slot.}%
\label{fig6}%
\vspace{-0.2in}
\end{figure}

\begin{figure}[t]%
\centering
    \includegraphics[scale=0.42]{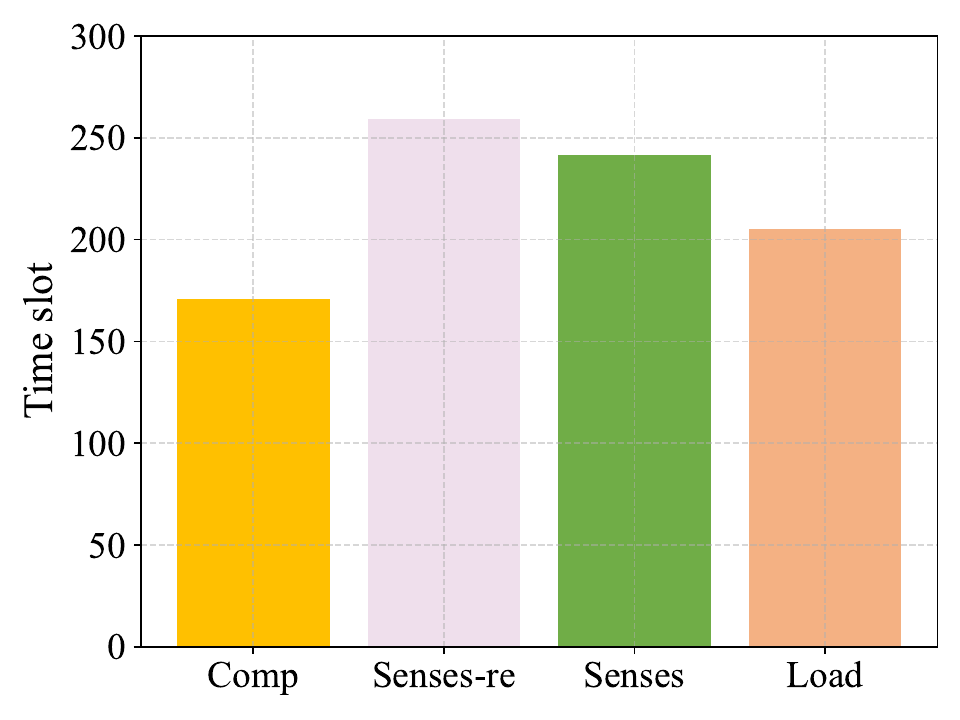}
\caption{Maximum operational duration under different energy-saving algorithms.}%
\label{fig7}%
\vspace{-0.2in}
\end{figure}

\subsubsection{Testbed} As shown in Fig. \ref{fig5}, to evaluate the performance of our Senses algorithm, we conducted comparative experiments in a real-world environment system that consists of $4$ Raspberry Pi devices (version $4$ $8G$) as edge servers and $6$ STM32F407ZG microcontrollers as IoT control devices. Each microcontroller is connected to a corresponding Raspberry Pi via Wi-Fi channels. Raspberry Pis are interconnected through Wi-Fi to form an edge computing environment. Although each microcontroller has computing and storage resources, they cannot support deploying complex optimization models. Therefore, we deploy Senses on edge servers. Sensors are integrated into each microcontroller for data collection. We model a heterogeneous IoT environment by turning off some of the microcontroller's resources. Then, we configure each microcontroller with a different capacity $5V$ batteries to support their operation, including three $3000mAh$ and three $2000mAh$ lithium batteries. All sensors are powered by microcontrollers. To facilitate measurement, a $20000$mAh lithium battery simultaneously powers $4$ Raspberry Pis and the power sensing modules to obtain each battery’s level. Considering that the loads on the microcontrollers are different, we use current sensors to determine their load conditions. We conduct multiple experiments to minimize the impact of sensor accuracy fluctuations and battery self-discharge on the experimental results. 

\subsubsection{Comparison methods}
To validate the performance of our algorithm, we have compared Senses against baselines. 

\begin{itemize}
\item[~] \textbf{Simulation}
\item \textit{Comp}, which de-duplicates the data by a hashing algorithm~\cite{chen2015compressing,zhao2020asymmetric}. This method de-duplicates and compresses the sensor data uploaded by the control device at the edge server.

\item \textit{Load}, which offloads compute tasks from the control device to the edge server~\cite{mao2016dynamic,tong2022dynamic}. This approach achieves energy-saving computing offloading based on the power consumption of the control device.

\item \textit{Senses-re}, which has the same workflow as Senses. However, the control device retains duplicate data for sensors with a fixed range.
\end{itemize}

\begin{itemize}
\item[~] \textbf{Testbed}
 
\item \textit{Base}. The critical difference between the Base and Senses is that the Base does not adaptively adjust the operating range of sensors based on their status and environment. For sensors with fixed ranges, Base retains redundant data.
\end{itemize}

\begin{figure}[t]%
\centering
    \includegraphics[scale=0.38]{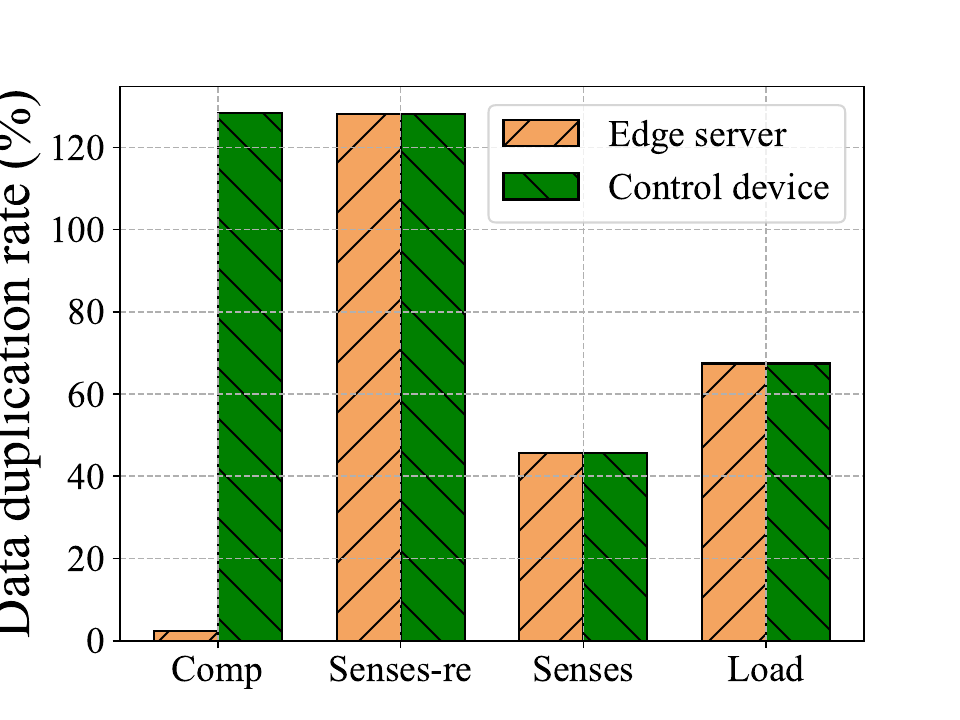}
\caption{Data duplication rate on edge servers and control devices.}%
\label{fig8}%

\end{figure}

\begin{figure}
\centering 
   
    \includegraphics[scale=0.33]{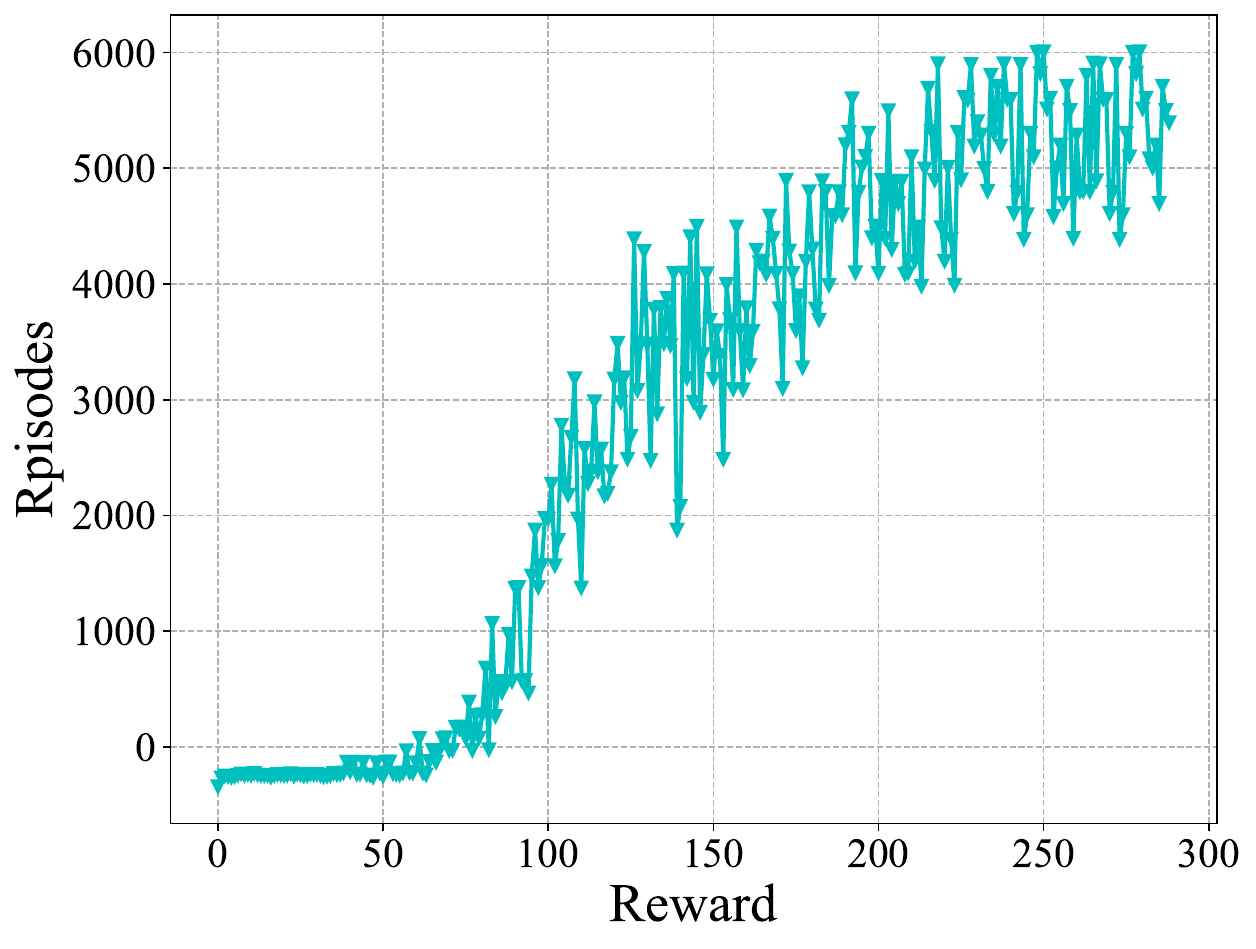}         
\caption{Training convergence of MARL agent with reward versus No. of episodes.} 
\label{fig9}  
\end{figure} 

\begin{figure}
\centering
	\subfloat[Senses.]{
		\label{fig10:1}     
		\includegraphics[scale = 0.28]{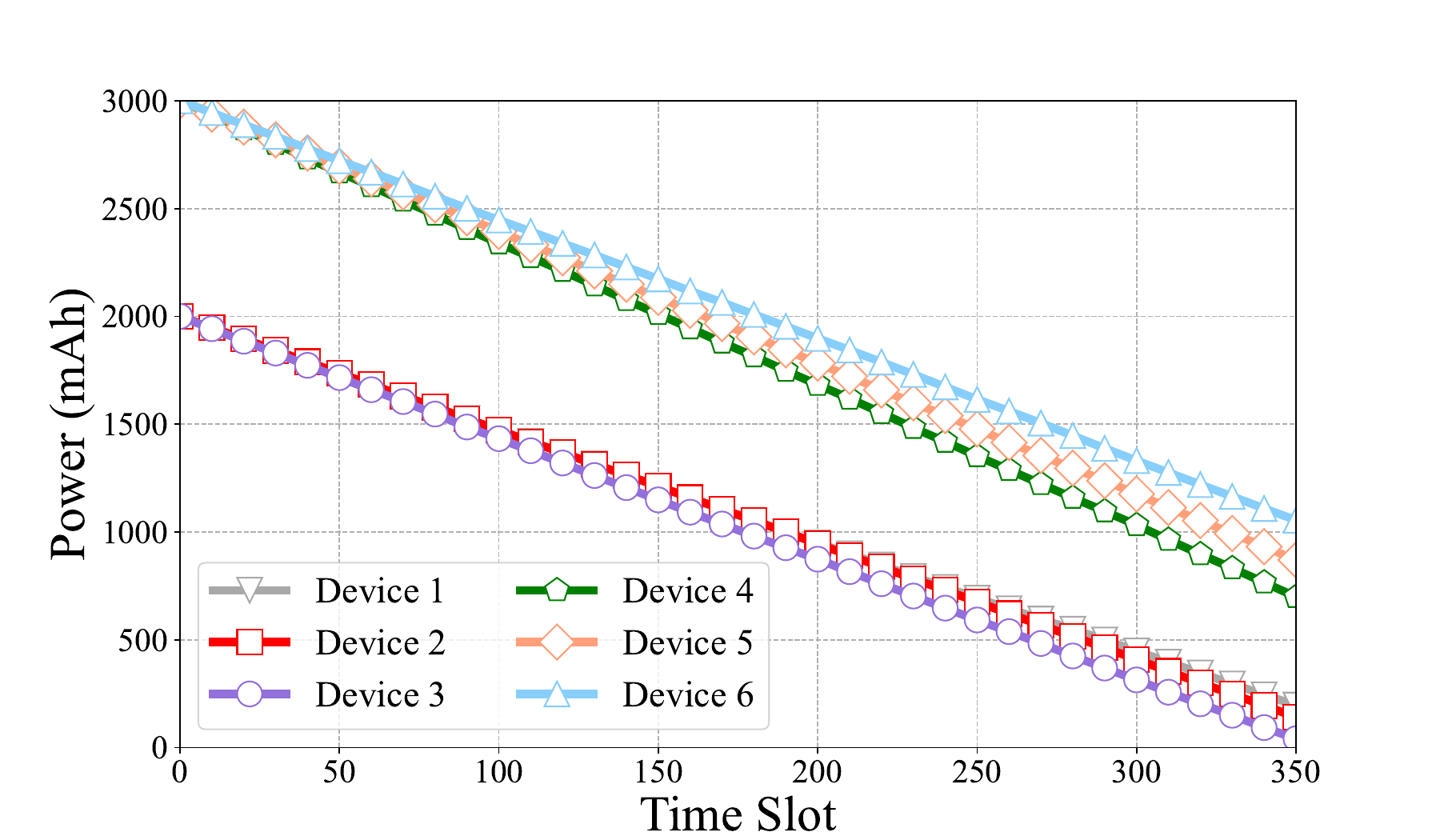} }
  
	\subfloat[Base.]{ 
		\label{fig10:2}     
		\includegraphics[scale = 0.28]{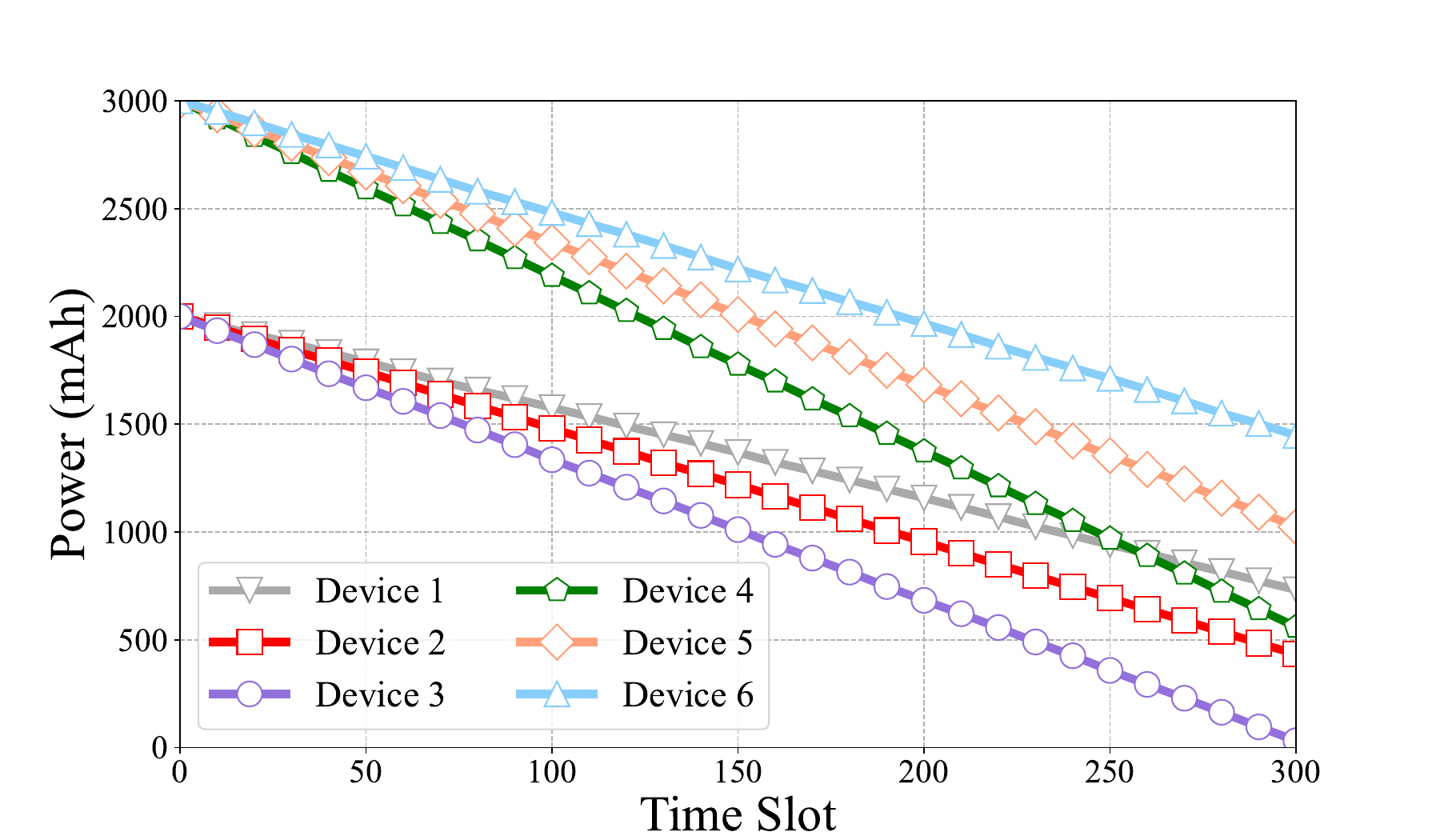}     
	}    
	\caption{Power consumption on control devices.}   
	\label{fig10} 

\end{figure} 

\subsection{Simulation Results}\label{sec5:2}
\subsubsection{Energy consumption over time} We select the control device as the comparison infrastructure. Compared to the edge server, it is more constrained in terms of energy. The experimental results are illustrated in Fig. \ref{fig6},
where different methods share the same settings such
as adjustable ranges and numbers of sensors in each set of experiments. Among all comparison methods, Senses demonstrates the lowest energy consumption, followed by Senses-re. Compared with Comv and Load, Senses reduces energy consumption by $30.49\%$ and $21.48\%$, respectively. Compared to Senses-re, which retains the duplicate data of sensors with fixed ranges, Senses still manages to reduce energy consumption by $8.84\%$.

\begin{figure}
\centering
	\subfloat[Senses.]{
		\label{fig11:1}     
		\includegraphics[scale = 0.28]{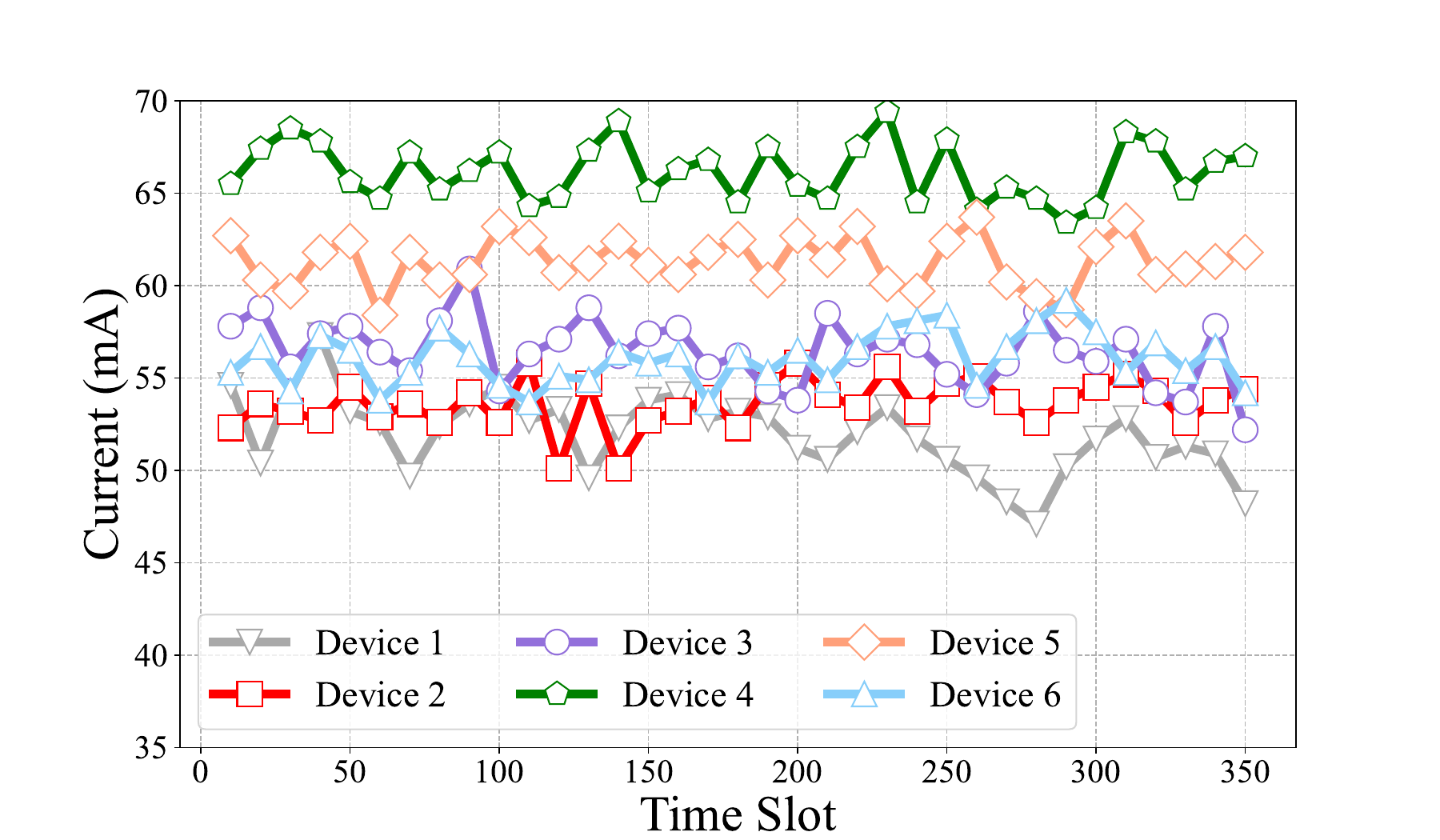} 
	}  
 
	\subfloat[Base.]{ 
		\label{fig11:2}     
		\includegraphics[scale = 0.28]{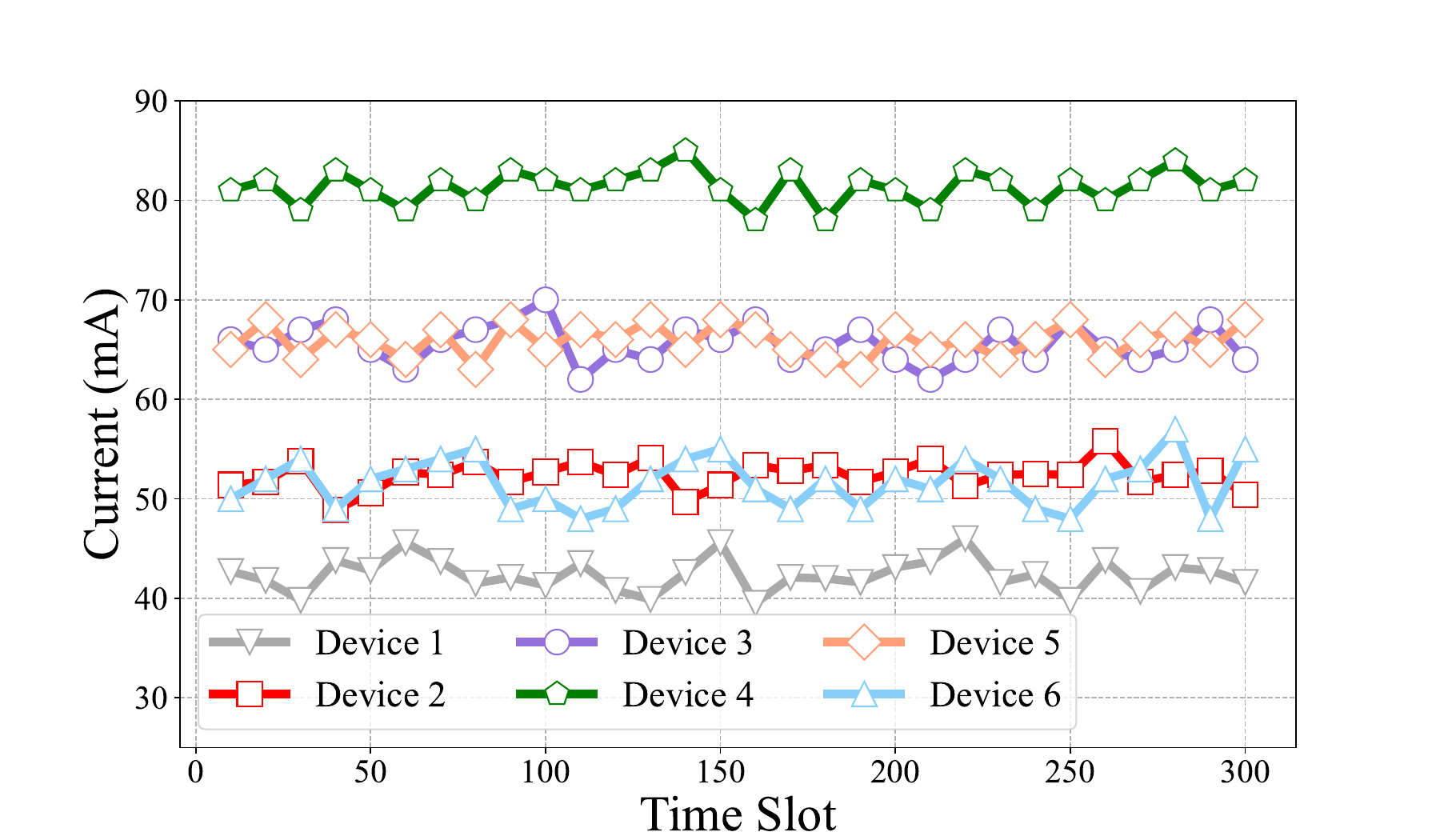}     

	}    
	\caption{Operating current on control devices.}   
	\label{fig11} 

\end{figure}

\begin{figure*}[]
\centering
	\subfloat{
		\label{fig12:1}     
		\includegraphics[scale=0.19]{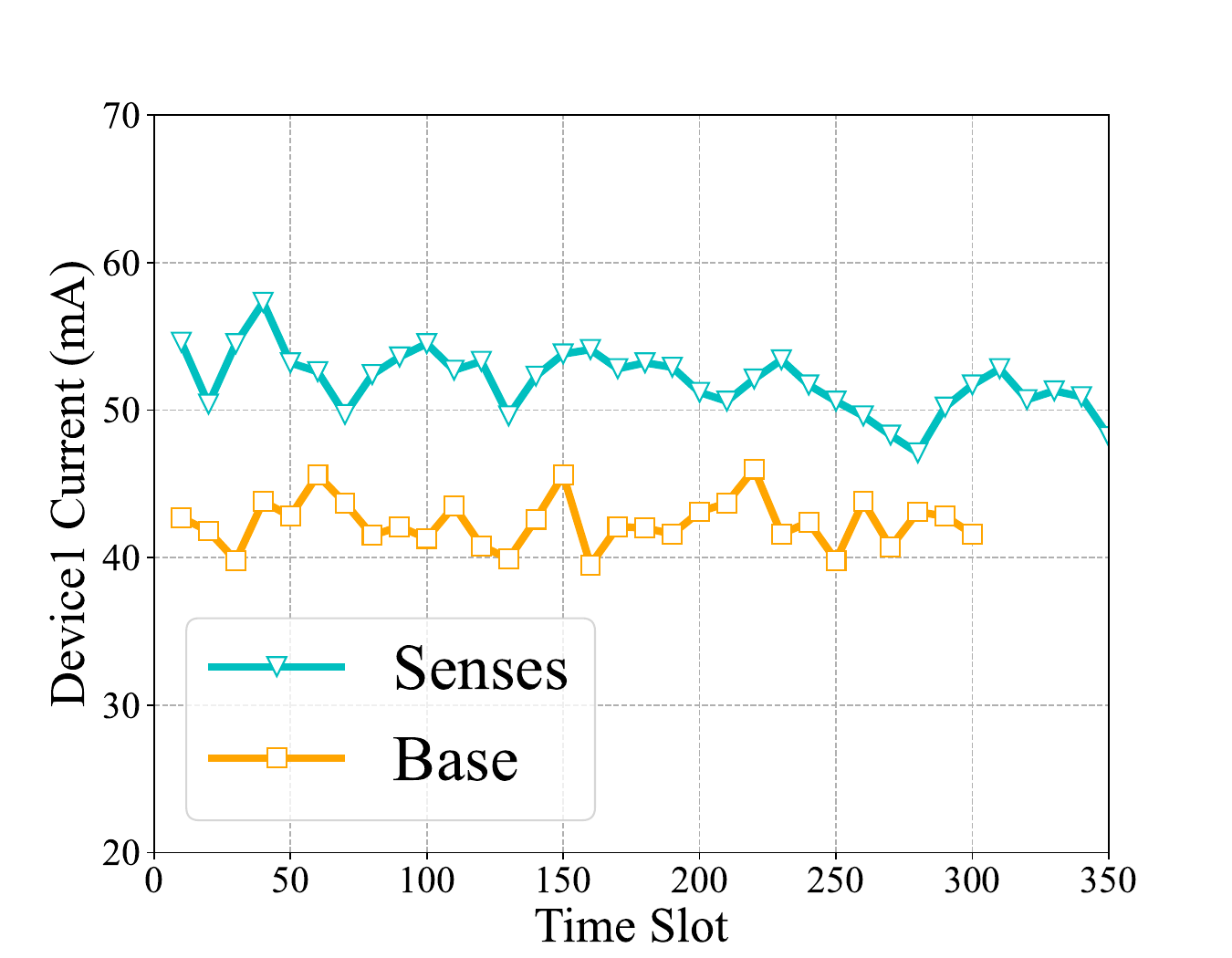} 
	}   
	\subfloat{ 
		\label{fig12:2}     
		\includegraphics[scale=0.19]{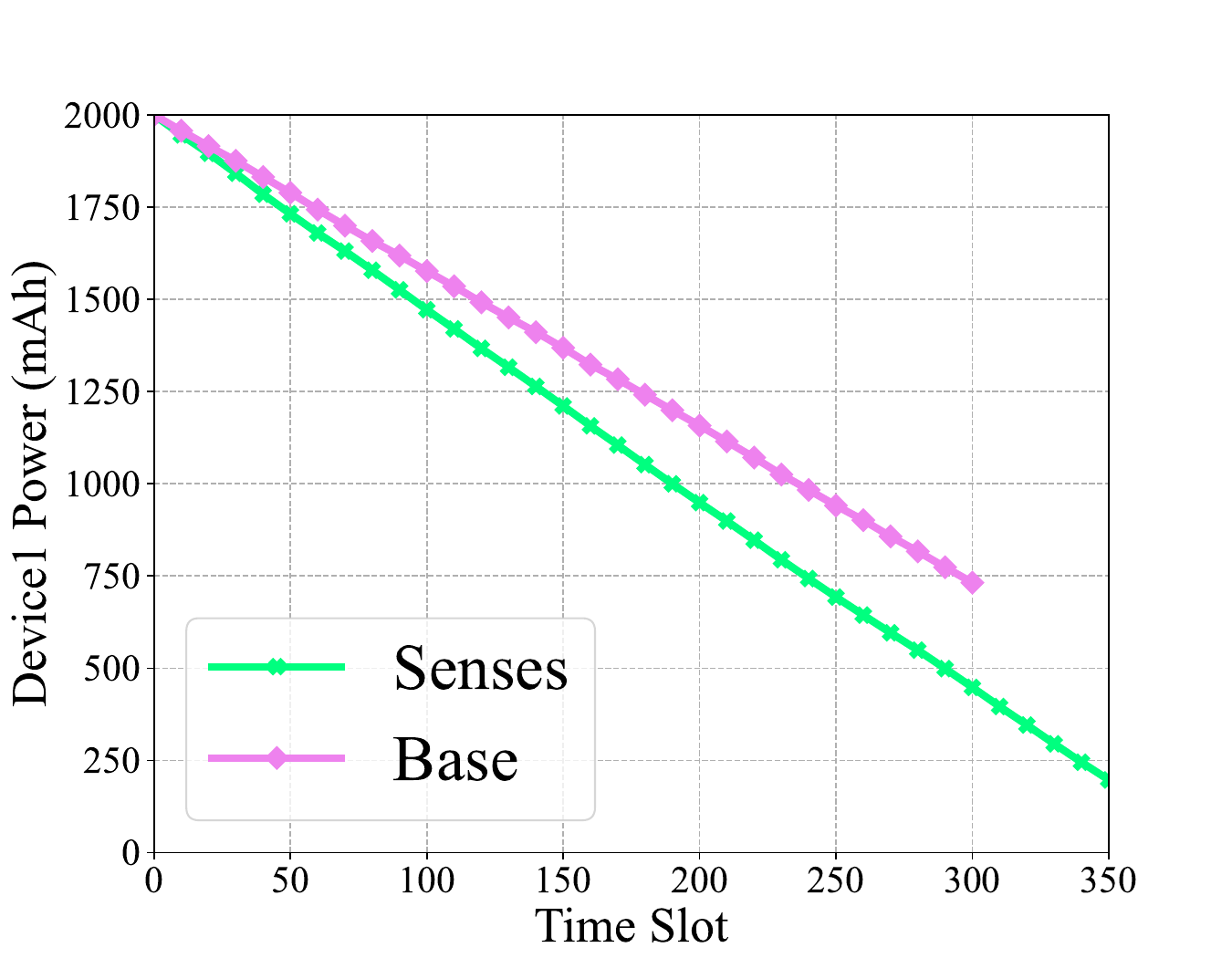}     

	}    
    \subfloat{
		\label{fig12:3}     
		\includegraphics[scale=0.19]{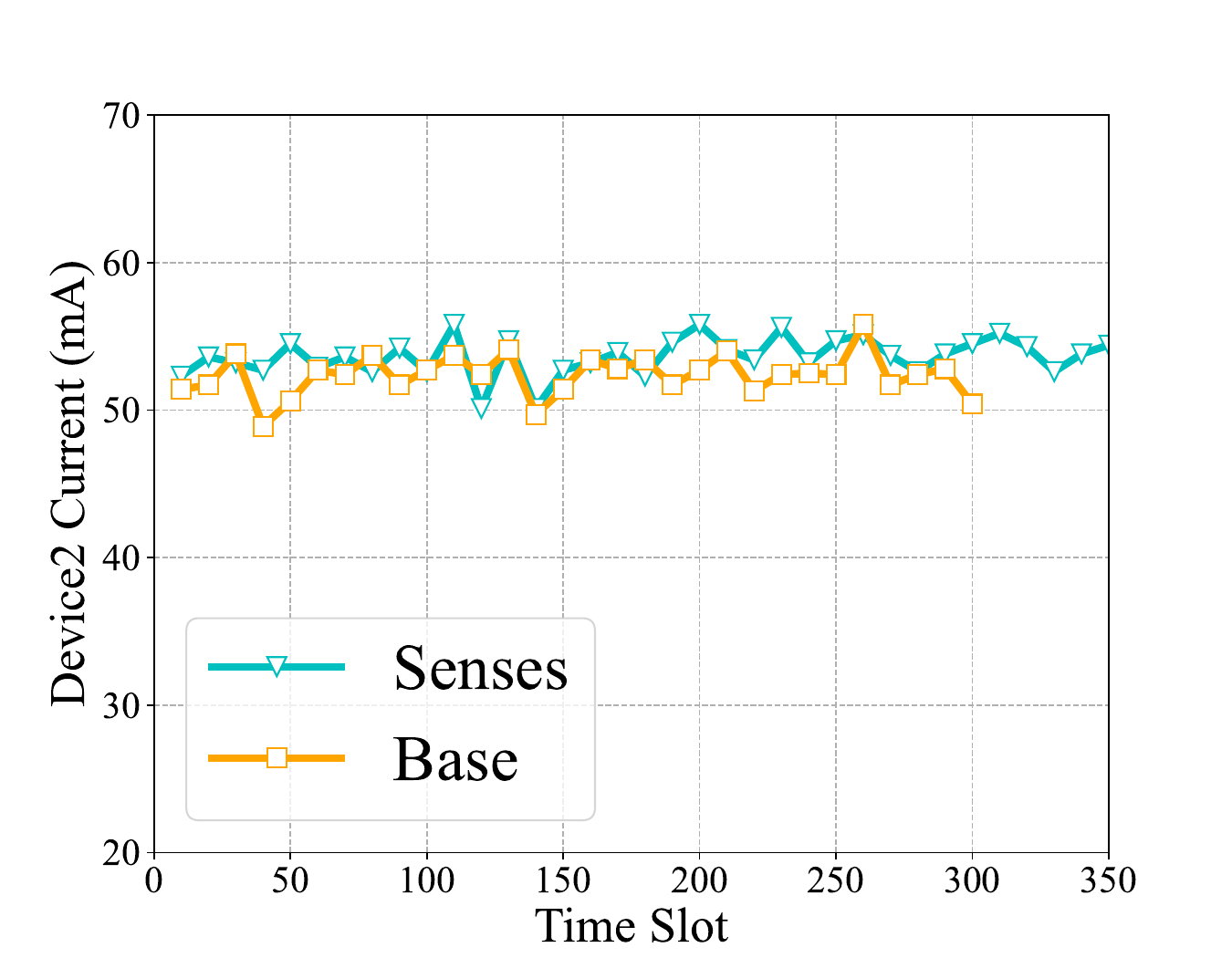} 
	}   
	\subfloat{ 
		\label{fig12:4}     
		\includegraphics[scale=0.19]{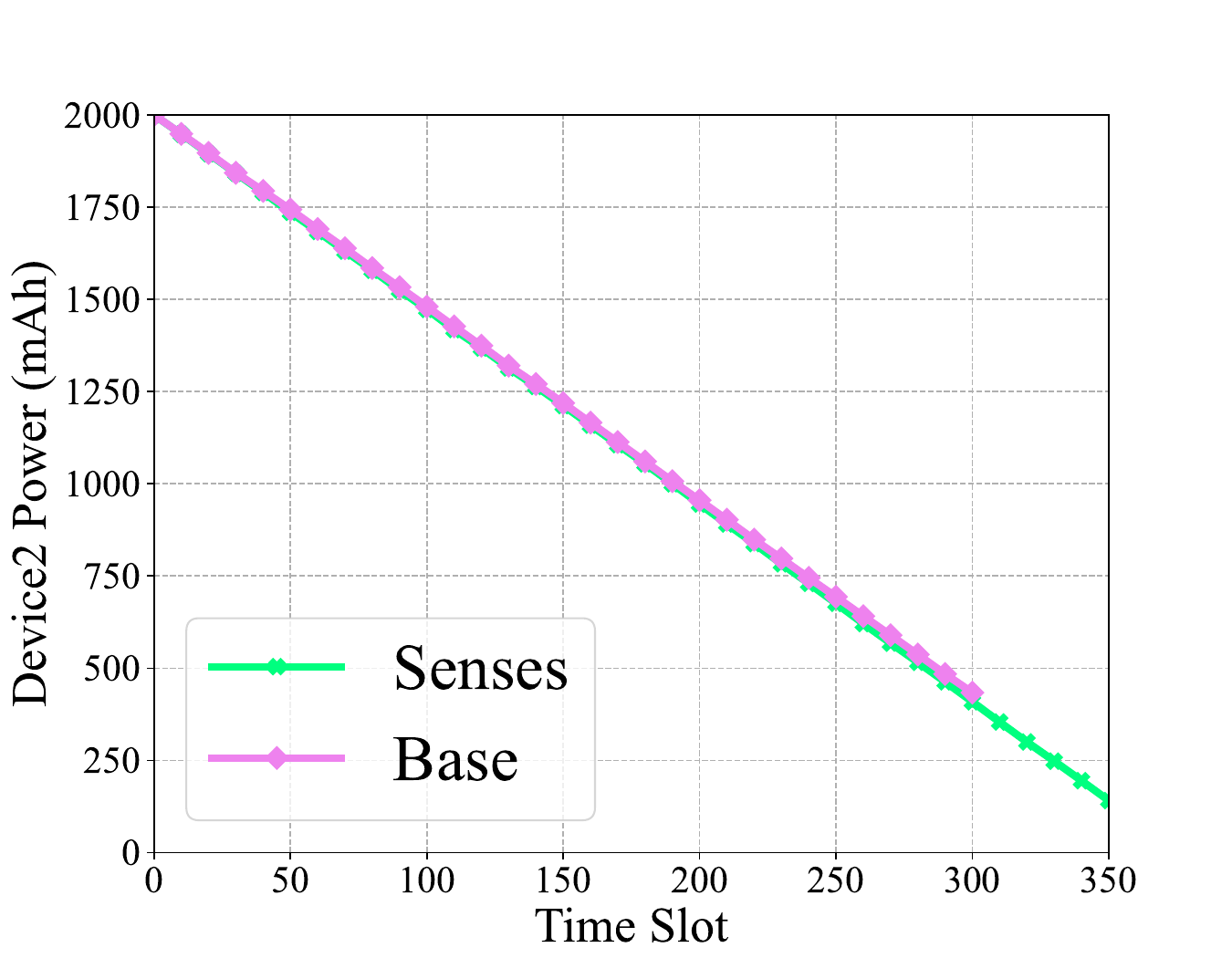}

	}  

	\subfloat{
		\label{fig12:5}     
		\includegraphics[scale=0.19]{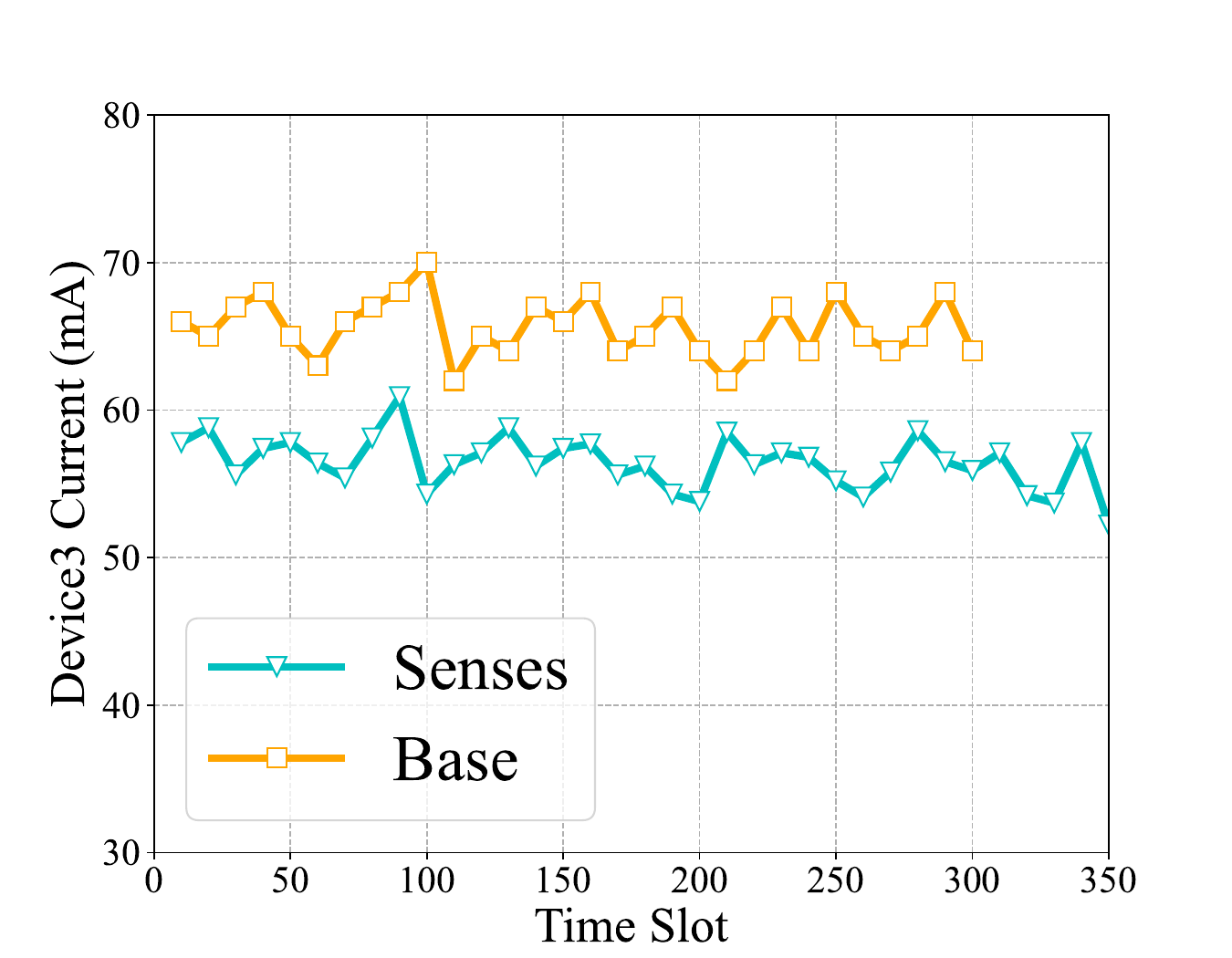} 
	}   
	\subfloat{ 
		\label{fig12:6}     
		\includegraphics[scale=0.19]{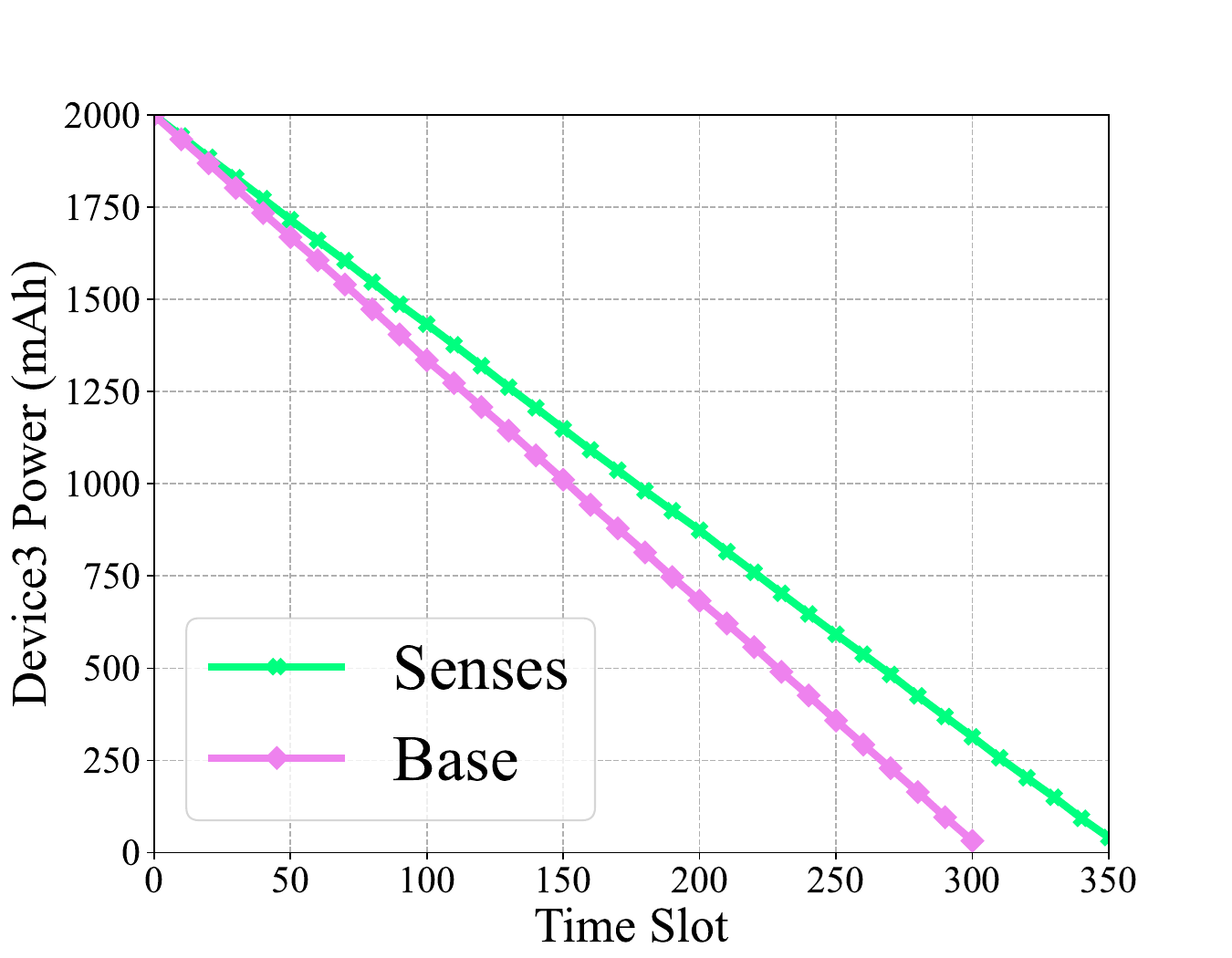}     

	}    
 \subfloat{
		\label{fig12:7}     
		\includegraphics[scale=0.19]{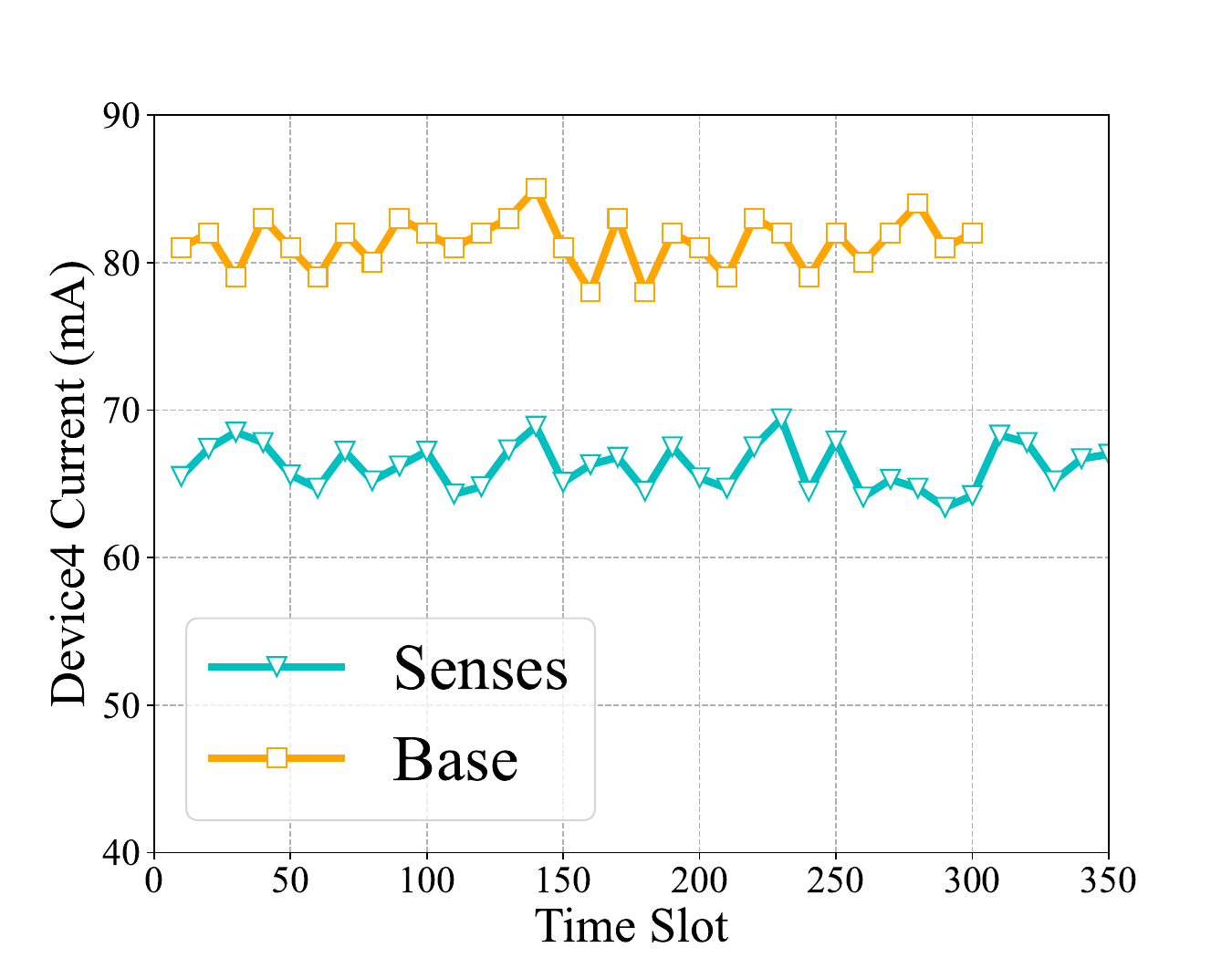} 
	}   
	\subfloat{ 
		\label{fig12:8}     
		\includegraphics[scale=0.19]{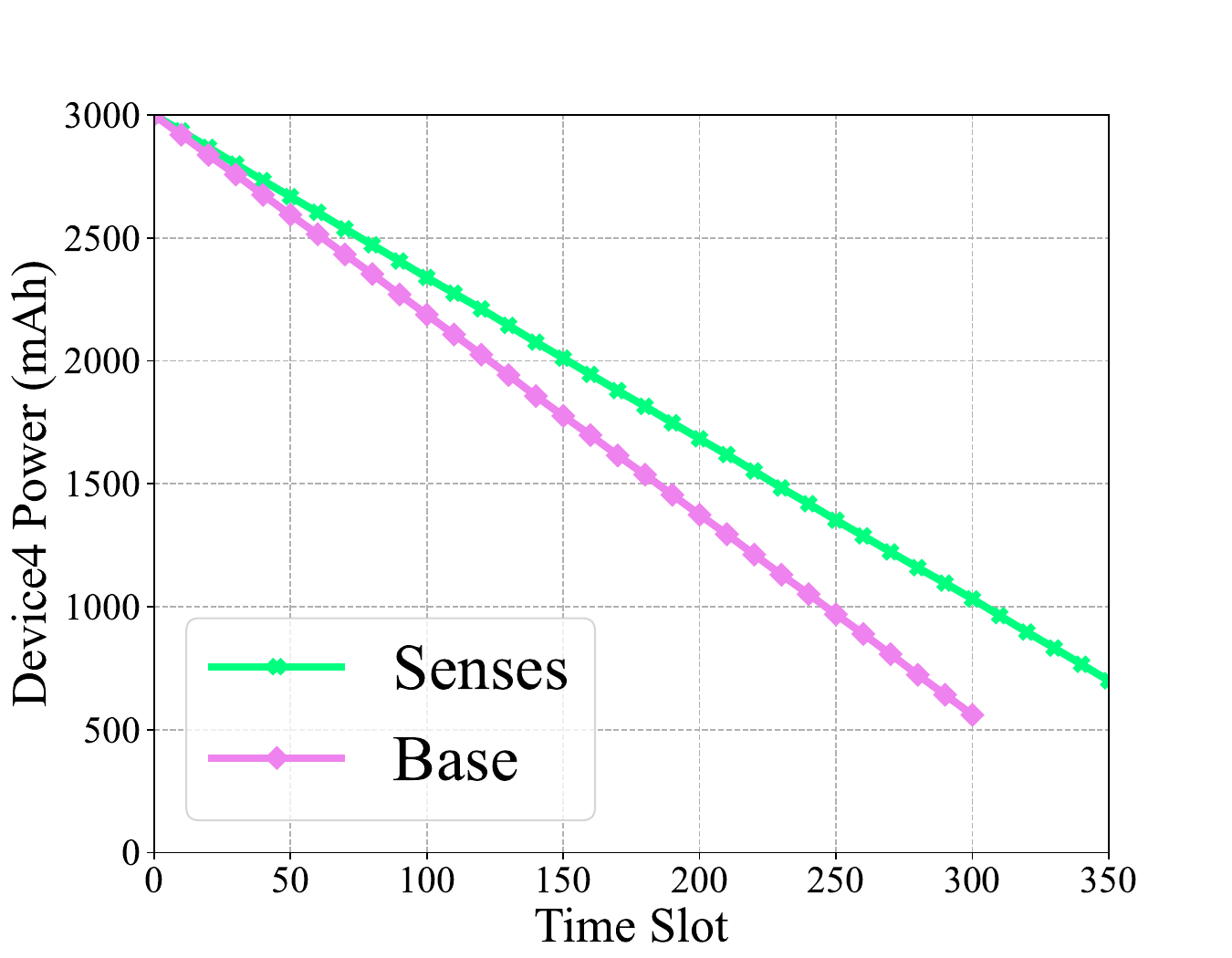}     

	}

 \subfloat{
		\label{fig12:9}     
		\includegraphics[scale=0.19]{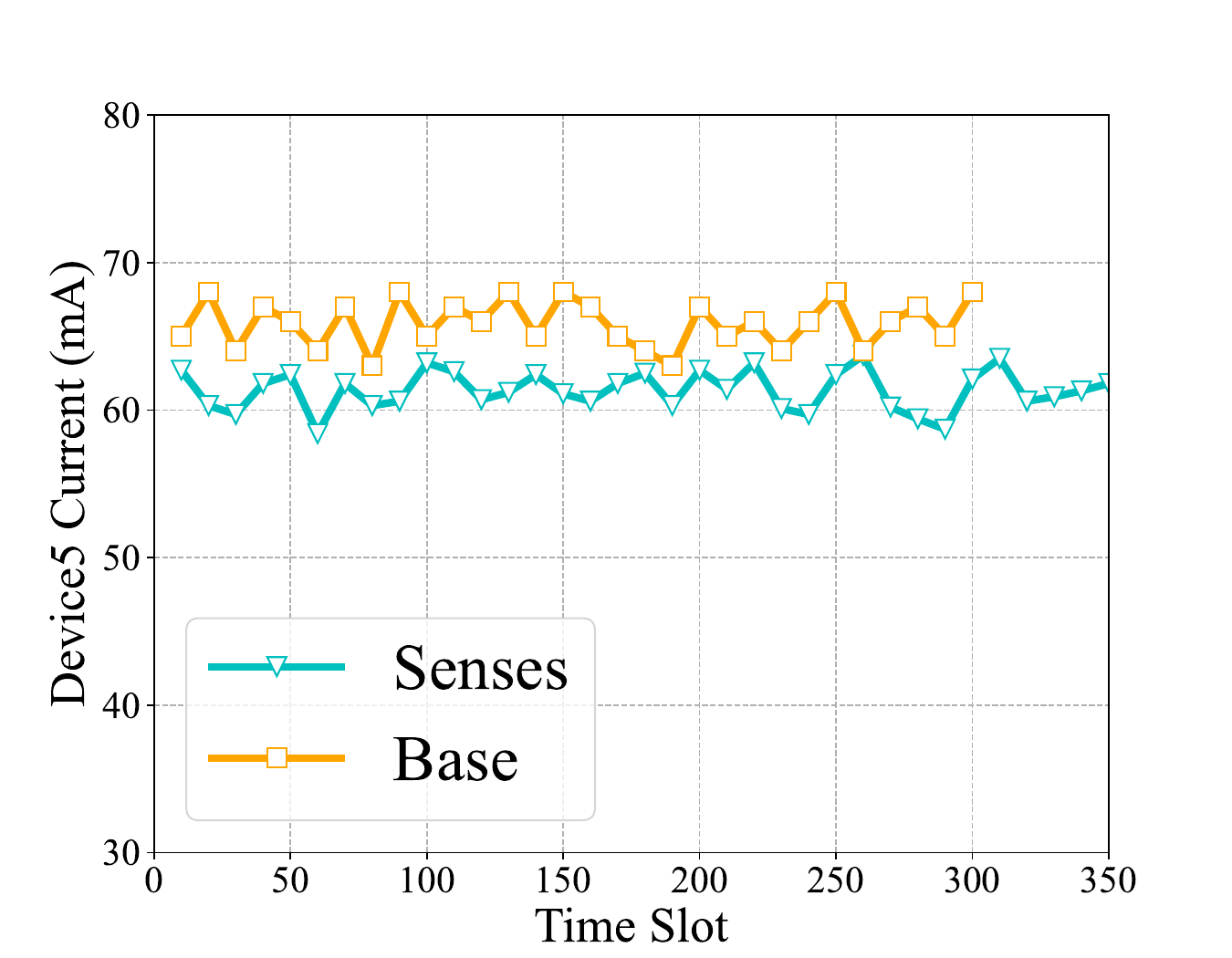} 
	}   
	\subfloat{ 
		\label{fig12:10}     
		\includegraphics[scale=0.19]{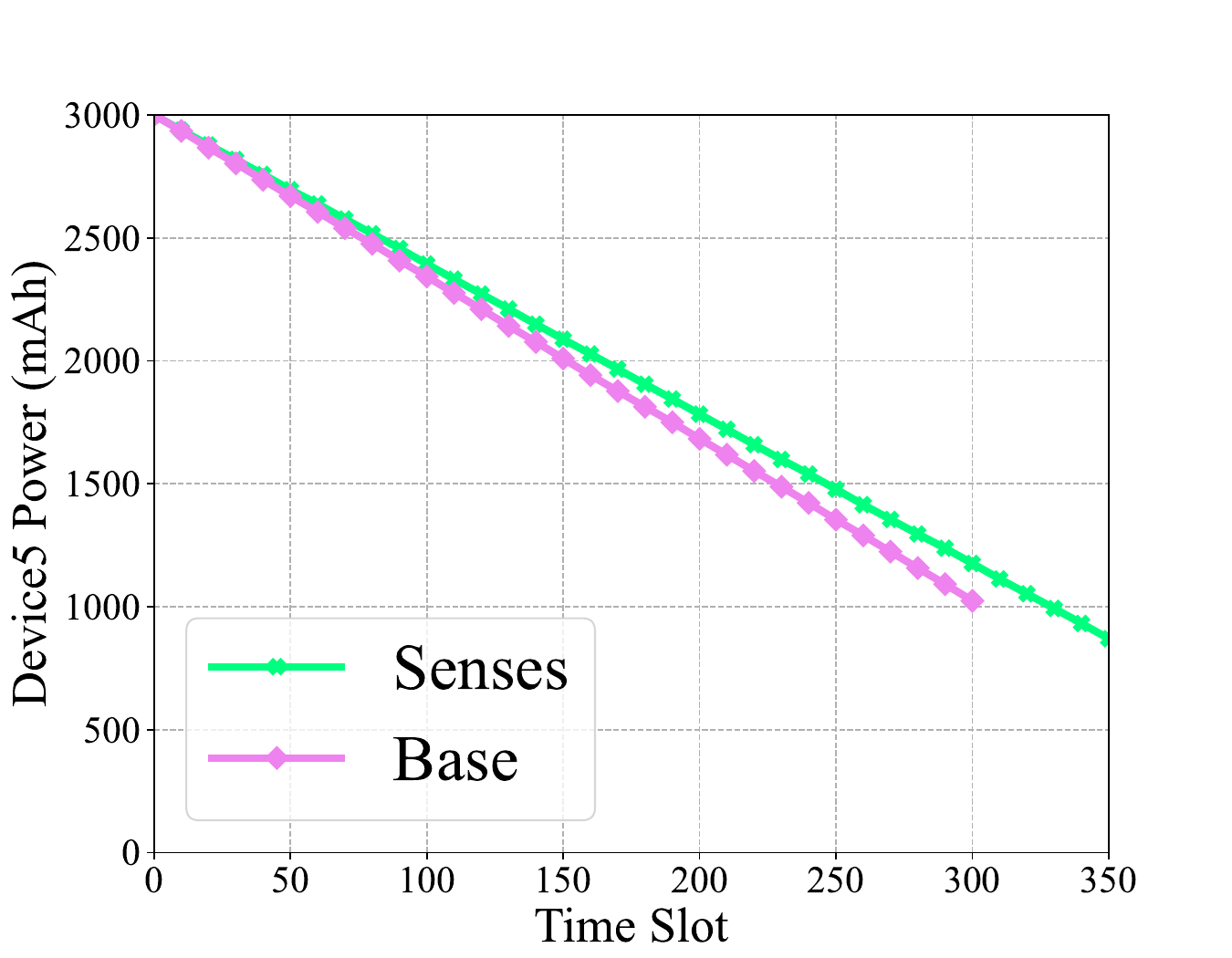}     

	}   
  \subfloat{
		\label{fig12:11}     
		\includegraphics[scale=0.19]{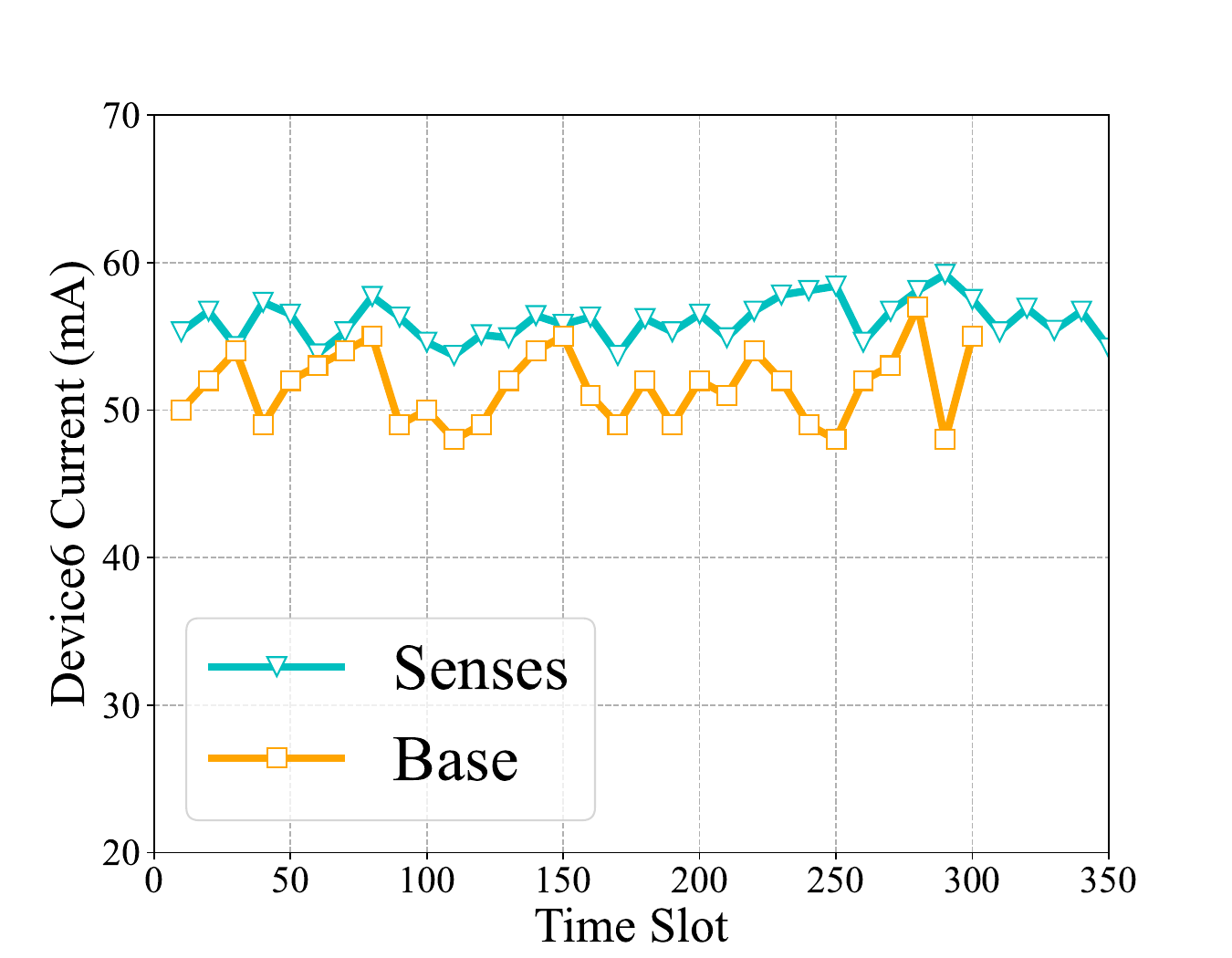} 
	}  
	\subfloat{ 
		\label{fig12:12}     
		\includegraphics[scale=0.19]{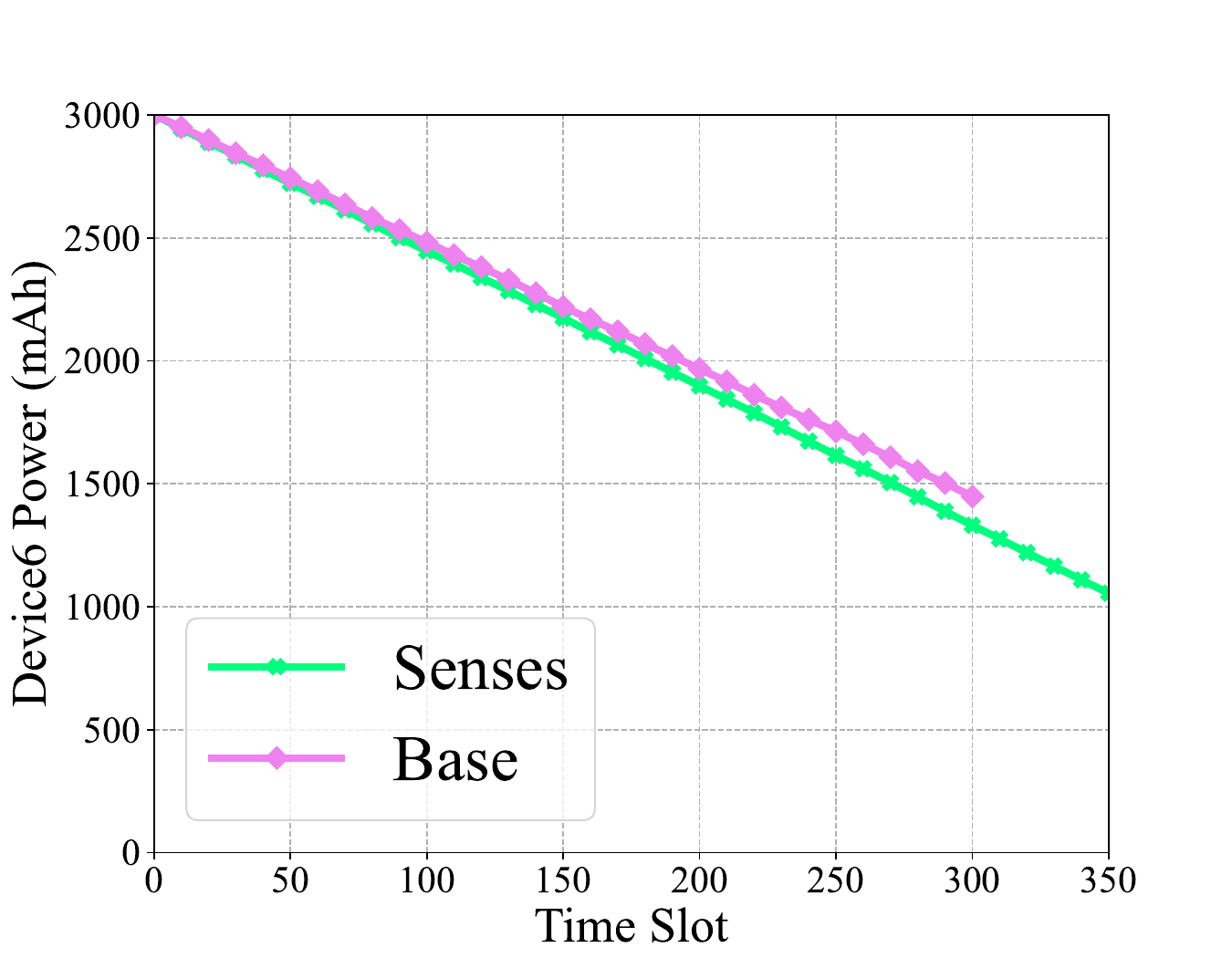}     

	}   
	\caption{Details of power consumption and operating current on each control device.}   
	\label{fig12} 

\end{figure*}

\subsubsection{Maximum operational duration}
When a control device depletes its energy, and no other control device is available to replace it to continue area data collection, we conclude that the system has reached its maximum operational duration. The maximum operational duration under different energy-saving algorithms is depicted in Fig. \ref{fig7}. Compared with Senses-re, Comv, and Load, Senses demonstrates the most extended operational duration and prolongs the maximum operational duration by $34.18\%$, $29.2\%$, and $16.77\%$, respectively.

\subsubsection{Data duplication rate}
Fig. \ref{fig8} illustrates the data duplication rates on edge servers and control devices for different energy-saving strategies. It is noteworthy that the Comv scheme exhibits remarkable performance on edge servers, followed by Senses. Although Senses shows a slightly lower improvement in Comv performance, it exhibits better energy-saving performance on control devices where energy is more scarce. In addition, Senses can be orthogonal with the current SOTA Comv and Load optimization techniques and act as the precursor step of
the existing methodologies. Therefore, Senses can be used as a plug-in and combined with them for energy saving.

\subsection{Testbed Results}\label{sec5:3}
\subsubsection{Training MARL Agent} Experience-driven MARL schemes can be divided into two stages: $(1)$ the training stage, which involves interacting with the environment and generating policies based on a reward function; and $(2)$ the online operation stage, which deploys policy on edge servers to determine sensor coverage. Edge servers have sufficient computational and storage resources compared to IoT control devices, so both stages of the MARL scheme can be completed on the edge servers.  MARL training was conducted in a system with $6$ IoT control devices. We observed how the training reward in MARL changed as the number of episodes increased. 
As shown in Fig. \ref{fig9}, during a MARL training process, the reward value increases consistently and becomes constant after $250$ episodes. 
This phenomenon indicates that the MARL model learns better policies and achieves better rewards.

\subsubsection{Energy Consumption} We initially test the power consumption of Senses and Base during their operation. We conduct ten trials for each indicator to mitigate the impacts of battery wear and current fluctuations. Fig. \ref{fig10} records the battery levels of $6$ control devices equipped with the base strategy. Devices $1\--3$ are furnished with $2000mAh$ batteries, while Devices $4\--6$ are outfitted with $3000mAh$ batteries. After $300$ time slots, Device $3$ lacks sufficient power to support the next time slot and ceases operation. Generally, the control device with the shortest operation time determines the overall system operational duration. Consequently, the overall operational duration of Base is $300$ time slots. However, devices operating under Senses still have sufficient charge after $300$ time slots and run out of power in the $350$ time slots. This is attributed to the fact that the adjustable sensor range strategy can transfer the power of Devices $4\--6$ to the depleted power Device $3$.
Compared with Base, Senses operates $20\%$ longer and has a narrower power gap between devices. In Senses, control devices with more power and lower loads can share the workload of control devices with insufficient power and higher loads under feasible circumstances. Thus, Senses exhibits better long-term availability in the IoT device network.

Fig. \ref{fig11:1} and Fig. \ref{fig11:2} quantify each device's current when Senses and Base are deployed. The current directly reflects the ongoing workload. As illustrated in Fig. \ref{fig11:2}, the operating currents of the $6$ devices are relatively dispersed, with a mean value of $59.7$ and a standard deviation of $12.82$. However, in Fig. \ref{fig11:1}, the operating currents of the $6$ devices are also relatively dispersed, having a mean value of $53.9$ and a standard deviation of $5.01$. This further showcases the superiority of Senses in dynamically equalizing sensor ranges and prolonging the overall operational duration.

\begin{figure}[t]
\centering
	
	\subfloat[Power consumption of all control devices.]{ 
		\label{fig13:1}     
		\includegraphics[scale = 0.28]{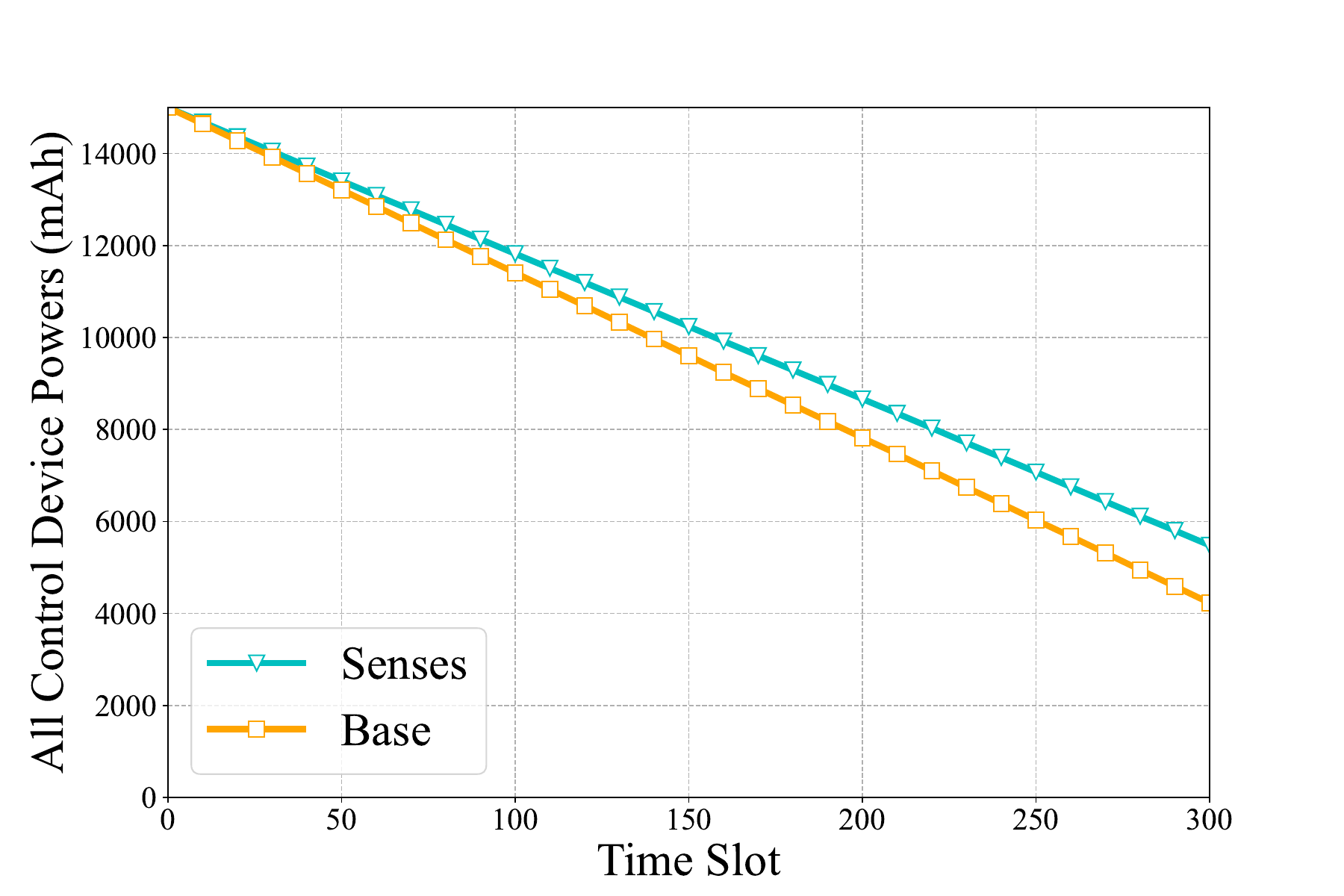}     

	} 
 
 \subfloat[Operating current of all control devices.]{
		\label{fig13:2}     
		\includegraphics[scale = 0.28]{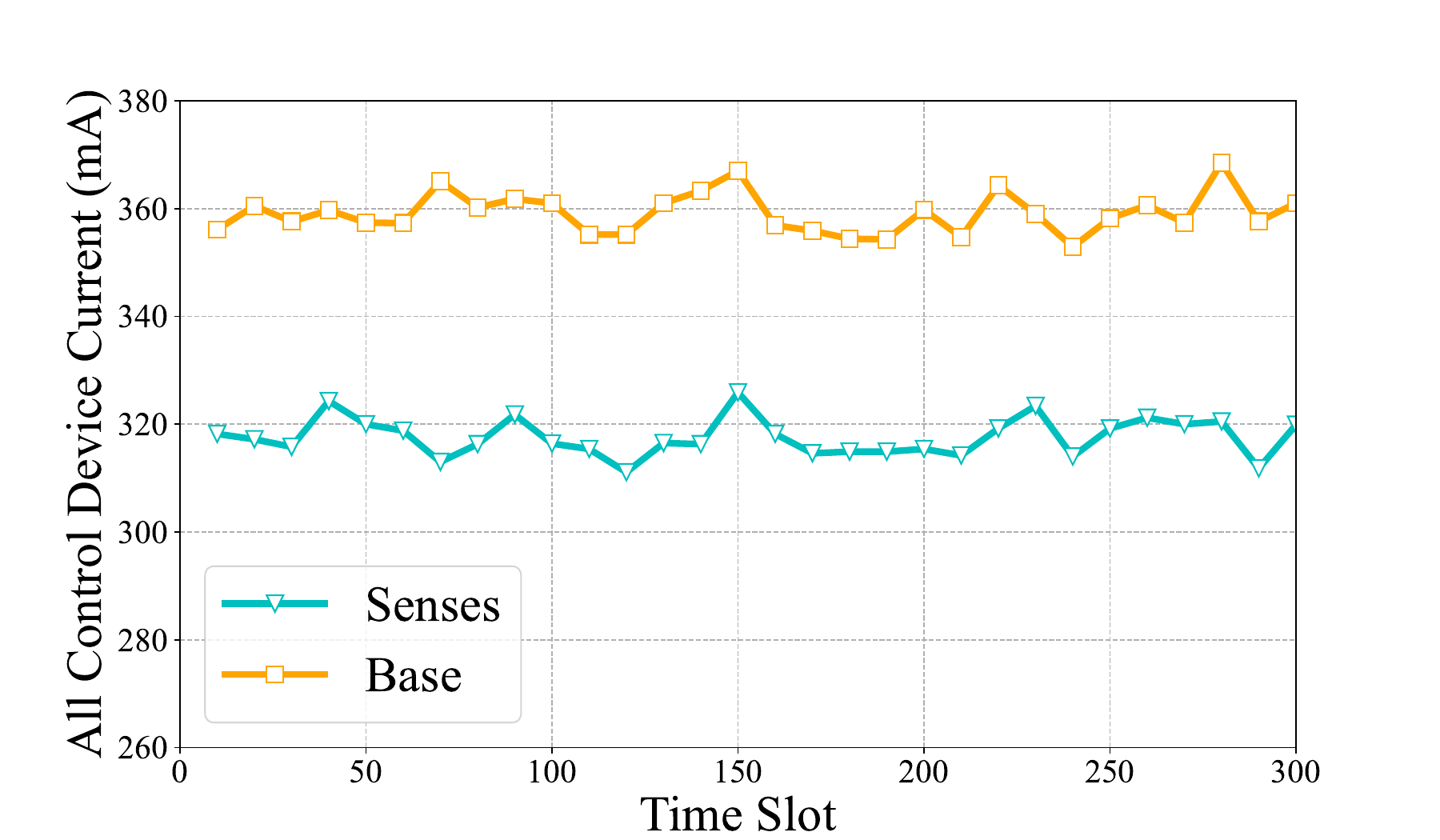} 
	}
	\caption{Total power consumption and operating current of control devices.}   
	\label{fig13} 

\end{figure} 

\begin{figure}[t]
\centering
	\subfloat[Operating current of edge servers.]{
		\label{fig14:1}     
		\includegraphics[scale = 0.28]{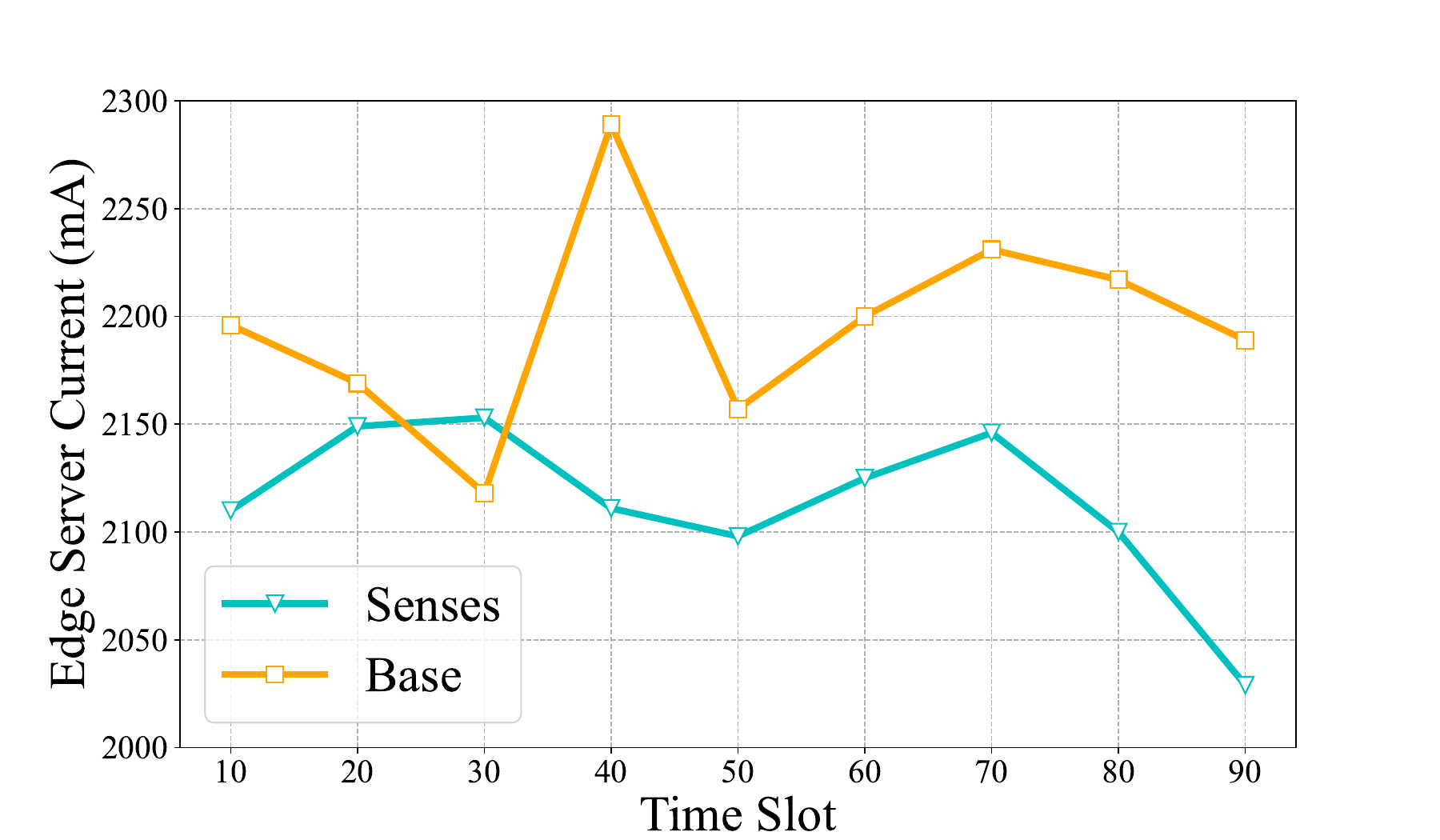} 
	}
 
	\subfloat[Power consumption of edge servers.]{ 
		\label{fig14:2}     
		\includegraphics[scale = 0.28]{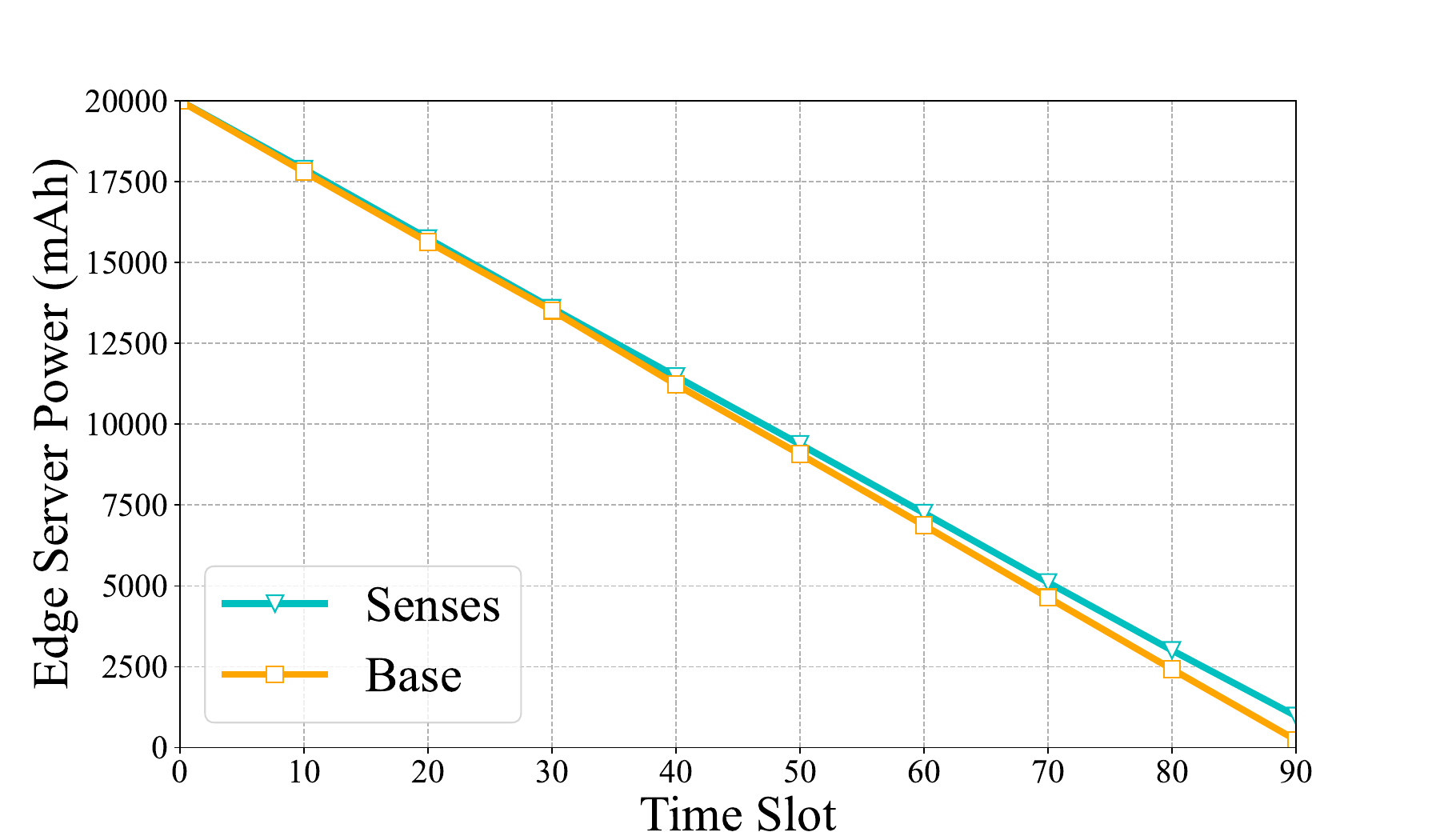}
  
	}    
	\caption{Total power consumption and operating current of edge servers.}   
	\label{fig14} 

\end{figure}

Fig. \ref{fig12} depicts the battery consumption and current for each control device in detail. Compared with Base, Senses exhibits similar current and power consumption on Device $2$, lower consumption on Devices $3\--5$, but higher consumption on Devices 1 and 6. In Senses, current and power consumption rises when one device shares the task of collecting data from another device. Meanwhile, that of the shared device turns low.

Fig. \ref{fig13} portrays the total energy consumption of IoT devices under Senses and Base within a cycle time. In Fig. \ref{fig13:1}, the total battery capacity of the $6$ IoT devices is $15,000$mAh. Unlike Base, Senses consumes less power over $300$ time slots, achieving an average energy savings of $11.37\%$ for IoT devices.
Fig. \ref{fig13:2} illustrates the total operating current of the $6$ IoT devices. Compared to Base, the Senses algorithm realizes an average current savings of $35$mA per time slot. Senses mainly contributes to reducing the data volume by adjusting the sensor range and deleting redundant data, thereby decreasing the energy consumption caused by data collection, transmission, and calculation. In the large-scale IoT, the energy-saving effect of Senses will be more prominent as the operational duration extends.

Subsequently, we conduct a similar experiment as that in Fig. \ref{fig13} on edge servers (Raspberry Pis). As depicted in Fig. \ref{fig14}, we compare the energy consumption and current draw caused by Senses and Base. Notably, Senses has only mediocre energy-saving advantages at the edge and, in some cases, may even lead to slightly higher energy consumption (e.g., at a time slot of $30$). This is because deploying Senses on the edge server generates more computational, storage, and communication overhead, which cannot be compensated by the energy savings achieved through reduced data volume.
On the other hand, considering that edge servers usually have stable and reliable power supplies and are unlikely to experience energy depletion, we believe that migrating part of the energy consumption from the control devices to edge servers is acceptable.

\section{Conclusion} \label{sec6}
This paper presents the Senses design in edge-IoT, with the ambition of better energy efficiency. At its core, Senses optimizes the volume of data so that energy consumption during transmission and computation can be reduced. For general scenarios with heterogeneous computing resources, energy, and workloads in intelligent control devices, Senses further balances the operational duration of control devices to prolong the overall operational duration of the IoT device network. Comprehensive experiments commit that Senses outperforms baselines in various experimental settings. It transmits/processes data with lower energy consumption in the same operational duration.

\section*{Acknowledgment}
The authors thank all the anonymous reviewers for the insightful feedback. Besides, this work is partially supported by National Natural Science Foundation of China under Grant No. $U23B2004$.

\bibliographystyle{IEEEtran}

\bibliography{sample-base}
%\bibliography{NFVmulticast}

\begin{IEEEbiography}[{\includegraphics[width=1in,height=1.25in, clip,keepaspectratio]{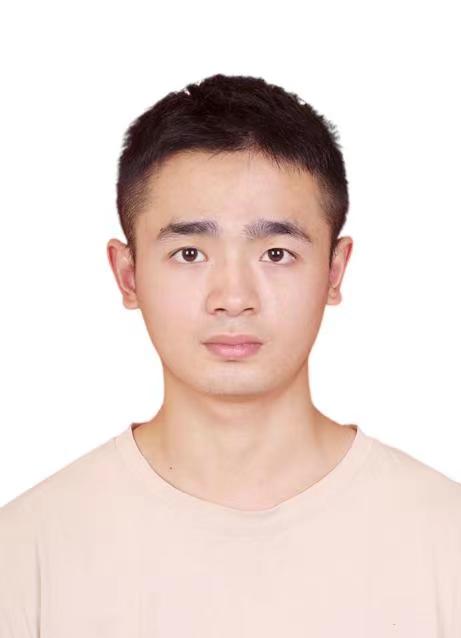}}]{Zongyang Yuan} received the B.S. degree in machinery, in 2021, Northeast Agricultural University in Harbin, China. Currently, he is a PhD student at College of Systems Engineering at National University of Defense Technology, China. His research interests include federated learning, internet of things, and edge computing.

\end{IEEEbiography}% \vspace{-0.35in}

\begin{IEEEbiography}[{\includegraphics[width=1in,height=1.25in,clip,keepaspectratio]{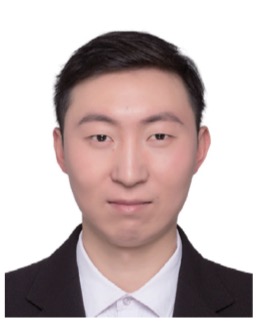}}]{Lailong Luo} eceived his B.S, M.S. and Ph.D. degree at the school of systems engineering from National University of Defense Technology, Changsha, China, in 2013, 2015 and 2019, respectively. He is currently a associate professor in the school of systems engineering, National University of Defense Technology, Changsha, China. His research interests include probabilistic data structures and data analysis.
\end{IEEEbiography}% \vspace{-0.35in}

\begin{IEEEbiography}[{\includegraphics[width=1in,height=1.25in,clip,keepaspectratio]{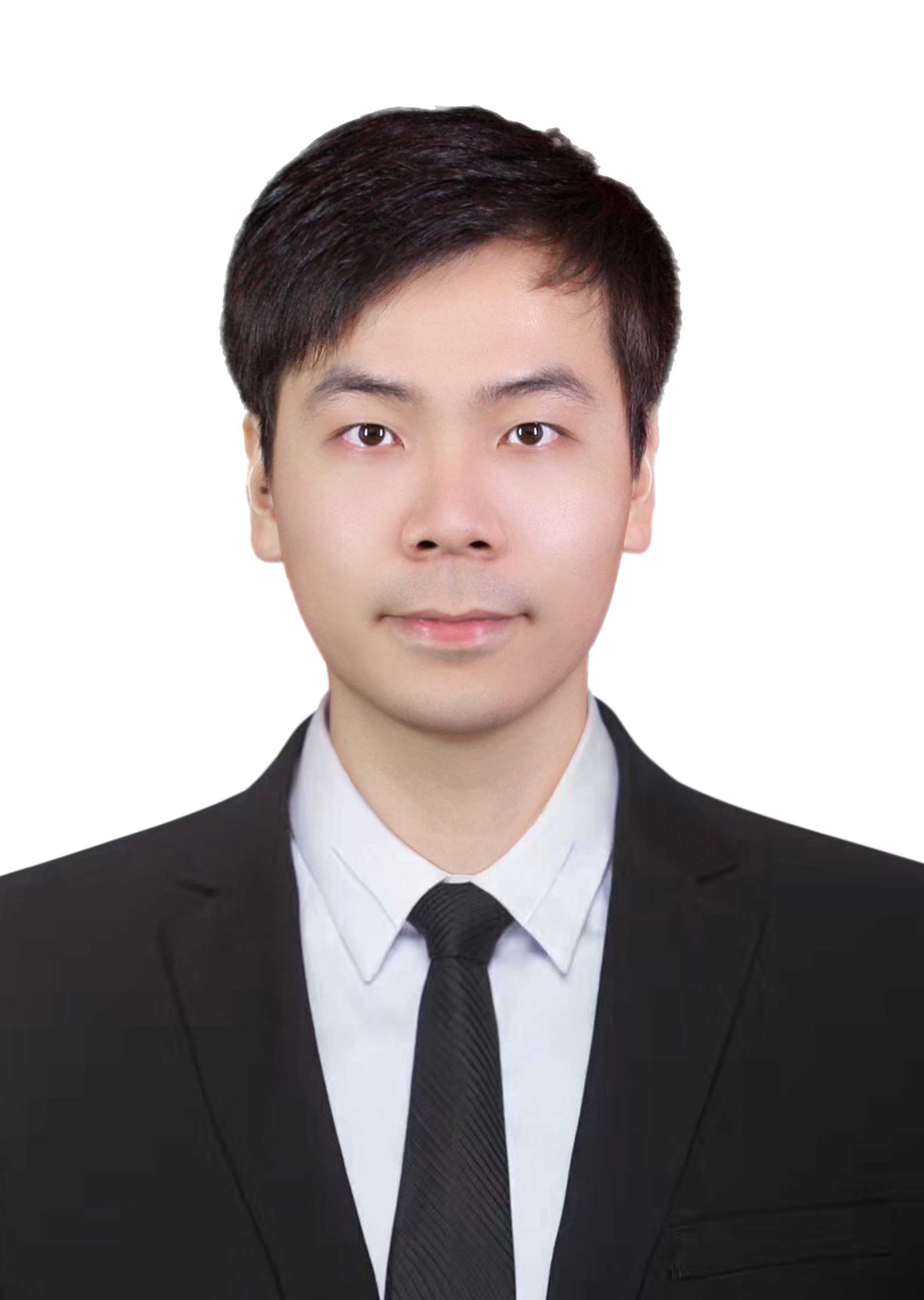}}]{Qianzhen Zhang} received his Ph.D. degree from National University of Defense Technology in 2022. Currently, he is a lecturer at College of Systems Engineering at National University of Defense Technology, China. His research interests include Continuous Subgraph Matching, Graph Data Analytics and Knowledge Graph.

\end{IEEEbiography}% \vspace{-0.35in}

\begin{IEEEbiography}[{\includegraphics[width=1in,height=1.25in,clip,keepaspectratio]{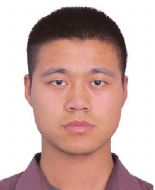}}]{Bangbang Ren} received his Ph.D. degree, M.S. degree and B.S. degree from the National University of Defense Technology, China, in 2021, 2017, and 2015, respectively. He was also a visiting research scholar of the University of Gottingen, German, in 2019. His research interests include software-defined network and network optimization.
\end{IEEEbiography}% \vspace{-0.35in}

\begin{IEEEbiography}[{\includegraphics[width=1in,height=1.25in,clip,keepaspectratio]{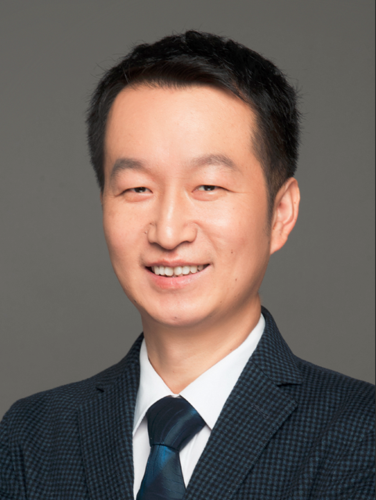}}]{Deke Guo} received the B.S. degree in industry engineering from the Beijing University of Aeronautics and Astronautics, Beijing, China, in 2001, and the Ph.D. degree in management science and engineering from the National University of Defense Technology, Changsha, China, in 2008. He is currently a Professor with the College of System Engineering, National University of Defense Technology. His research interests include distributed systems, software defined networking, data center networking, wireless and mobile systems, and interconnection networks.
\end{IEEEbiography}% \vspace{-0.35in}

\begin{IEEEbiography}[{\includegraphics[width=1in,height=1.25in,clip,keepaspectratio]{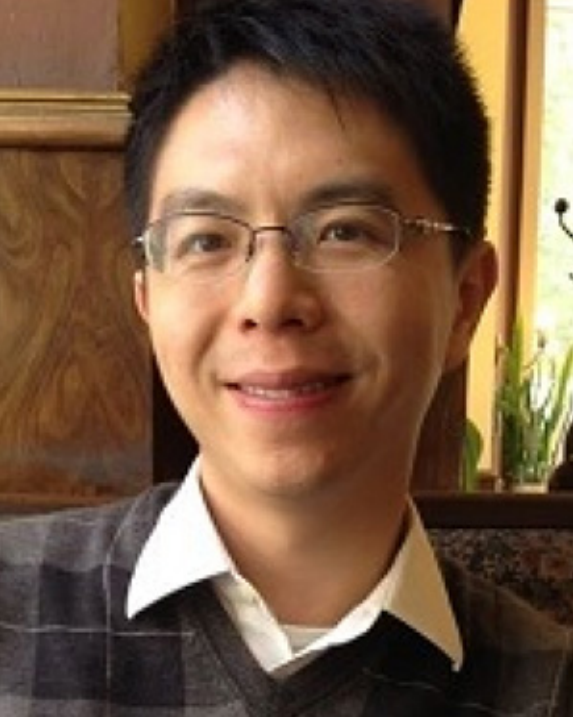}}]
{Richard T. B. Ma} (Senior Member, IEEE) received the B.Sc. degree (Hons.) in computer science and the M.Phil. degree in computer science and engineering
from The Chinese University of Hong Kong, in 2002 and 2004, respectively, and the Ph.D. degree in electrical engineering from Columbia University in
2010. During his Ph.D. degree, he was a Research Intern with IBM T. J. Watson Research Center, NY, USA, and Telefonica Research, Barcelona. He is currently
an Associate Professor with the Department of Computer Science, National University of Singapore. His current research interests include distributed systems and network economics. He is a co-recipient of the Best Paper Award in the IEEE Workshop on Smart Data Pricing 2015, the IEEE ICNP 2014, and the
IEEE IC2E 2013.
\end{IEEEbiography}% \vspace{-0.35in}

\vfill
\end{document}